\documentclass[12pt]{article}

\pdfoutput=1

\usepackage{colonequals}
\usepackage{color}
\usepackage{dsfont}
\usepackage{epsfig, palatino}
\usepackage{pstricks,pst-node,pst-tree}
\usepackage{epic}
\usepackage{mathrsfs}
\usepackage{ae} 
\usepackage[T1]{fontenc}
\usepackage[ansinew]{inputenc}
\usepackage{amsmath}
\usepackage{amssymb}
\usepackage{graphicx}
\usepackage{color}
\definecolor{darkblue}{cmyk}{0.9,0.9,0,0}
\usepackage[colorlinks=true,linkcolor=darkblue,citecolor=darkblue,urlcolor=darkblue]{hyperref}
\usepackage{cite}
\usepackage{hyperref}
\usepackage{wasysym}
\usepackage{varioref}
\usepackage{makeidx}
\usepackage[english]{babel}
\usepackage{simplewick}
\usepackage{array}
\usepackage{multirow}

\usepackage{animate}
\usepackage[font={small}]{caption}

\renewcommand\]{\end{equation}}
\renewcommand\[{\begin{equation}}
\newcommand{\reef}[1]{(\ref{#1})}

\newcommand{\cO}{\mathcal O}

\newcommand{\D}{\Delta}

\newcommand{\be}{\begin{equation}}
\newcommand{\ee}{\end{equation}}
\newcommand{\bea}{\begin{eqnarray}}
\newcommand{\eea}{\end{eqnarray}}

\newcommand{\comment}[1]{}

\newcommand{\beq}{\begin{equation}}
\newcommand{\eeq}{\end{equation}}
\newcommand{\beqq}{\begin{equation*}}
\newcommand{\eeqq}{\end{equation*}}
\newcommand\beqa{\begin{eqnarray}}
\newcommand\eeqa{\end{eqnarray}}
\newcommand\beqaa{\begin{eqnarray*}}
\newcommand\eeqaa{\end{eqnarray*}}

\newcommand{\neqa}{\nonumber\end{eqnarray}} 
\newcommand{\la}[1]{\label{#1}}

\renewcommand{\d}{\partial}

\newcommand{\<}{{\langle}}
\renewcommand{\>}{{\rangle}}

\newcommand{\re}{\relax{\rm I\kern-.18em R}}

\renewcommand{\sp}{p\hspace{-.40em}/}

\definecolor{darkgreen}{rgb}{0.0, 0.45, 0.0}

\def\XXint#1#2#3{{\setbox0=\hbox{$#1{#2#3}{\int}$}
\vcenter{\hbox{$#2#3$}}\kern-.5\wd0}}

\def\su2{{SU(2)}}

\def\({\left(}
\def\){\right)}

\def\<{\langle}
\def\>{\rangle}

\def\i2{\frac{i}{2}}

\def\spi{\relax{\rm \pi\kern-0.5em /}}
\def\sA{\relax{\rm A\kern-0.5em /}}
\def\sp{\relax{\rm p\kern-0.5em /}}
\def\sd{\relax{\rm \d\kern-0.5em /}}
\def\sk{\relax{\rm k\kern-0.5em /}}
\def\sn{\relax{\rm n\kern-0.5em /}}
\def\sl{\relax{\rm l\kern-0.5em /}}
\def\sP{\relax{\rm P\kern-0.7em /}}
\def\sBethe{\relax{\rm \Bethe\kern-0.5em /}}

\def\cO{{\cal O}}

\def\2F1{\,_2{\rm F}_1}

\makeindex

\newcommand\blfootnote[1]{%
  \begingroup
  \renewcommand\thefootnote{}\footnote{\hspace{-6mm}#1}%
  \addtocounter{footnote}{-1}%
  \endgroup
}

\begin{document}

\thispagestyle{empty}

\renewcommand{\thefootnote}{\fnsymbol{footnote}}
\setcounter{page}{1}
\setcounter{footnote}{0}
\setcounter{figure}{0}

%%%%%%%%%%%%%%%%%%%%%%%%%%%%%%%%%%%%%%%%%%%%%%%%%%%%%%%%%%%%%%%%%%%%%%%%%%%%%%%%%%%%%%%%%%%%%%

\begin{flushright}
CERN-TH-2016-162
\end{flushright}
\vspace{-0.4in}
\begin{center}
$$$$
{\Large\textbf{\mathversion{bold}
The S-matrix Bootstrap I:\\QFT in AdS
}\par}
\vspace{1.0cm}

\textrm{Miguel F. Paulos$^\text{\tiny 1}$, Joao Penedones$^\text{\tiny 2,\tiny 3}$, Jonathan Toledo$^\text{\tiny 4}$, Balt C. van Rees$^\text{\tiny 5}$, Pedro Vieira$^\text{\tiny 4,\tiny 6}$}
\blfootnote{\tt  \#@gmail.com\&/@\{miguel.paulos,jpenedones,jonathan.campbell.toledo,baltvanrees,pedrogvieira\}}
\\ \vspace{1.2cm}
\footnotesize{\textit{
$^\text{\tiny 1}$Theoretical Physics Department, CERN, Geneva, Switzerland\\
$^\text{\tiny 2}$Institute of Physics, \'Ecole Polytechnique F\'ed\'erale de Lausanne, CH-1015 Lausanne,
Switzerland\\
$^\text{\tiny 3}$Centro de F\'isica do Porto,  Departamento de F\'isica e Astronomia,\\
Faculdade de Ci\^encias da Universidade do Porto,
Rua do Campo Alegre 687, 4169-007 Porto, Portugal\\
$^\text{\tiny 4}$Perimeter Institute for Theoretical Physics,
Waterloo, Ontario N2L 2Y5, Canada\\
$^\text{\tiny 5}$Centre for Particle Theory, Department of Mathematical Sciences, Durham University, Lower Mountjoy, Stockton Road, Durham, England, DH1 3LE\\
$^\text{\tiny 6}$ICTP South American Institute for Fundamental Research, IFT-UNESP, S\~ao Paulo, SP Brazil 01440-070}  
\vspace{4mm}
}

\par\vspace{1.5cm}

\textbf{Abstract}\vspace{2mm}
\end{center}
We propose a strategy to study massive Quantum Field Theory (QFT) using conformal bootstrap methods. The idea is to consider QFT in hyperbolic space and study correlation functions of its boundary operators. We show that these are solutions of the crossing equations in one lower dimension. By sending the curvature radius of the background hyperbolic space to infinity we expect to recover flat-space physics. We explain that this regime corresponds to large scaling dimensions of the boundary operators, and discuss how to obtain the flat-space scattering amplitudes from the corresponding limit of the boundary correlators. We implement this strategy to obtain universal bounds on the strength of cubic couplings in 2D flat-space QFTs using 1D conformal bootstrap techniques. Our numerical results match precisely the analytic bounds obtained in our companion paper using S-matrix bootstrap techniques.

\noindent

\setcounter{page}{1}
\renewcommand{\thefootnote}{\arabic{footnote}}
\setcounter{footnote}{0}

\setcounter{tocdepth}{2}

 \def\nref#1{{(\ref{#1})}}

\newpage

\tableofcontents

\parskip 5pt plus 1pt   \jot = 1.5ex

%%%%%%%%%%%%%%%%%%%%%%%%%%%%%%%%%%%%%%%%%%%%%%%%%%%%%%%%%%%%%%%%%%%%%%%%%%%%%%%%%%%%%%%%%%%%%%%%%%%
\newpage
\section{Introduction} 
In conformal invariant field theories, 
the correlation functions
of local operators are strongly constrained by virtue of the operator-state
correspondence, which results in a convergent operator product expansion
and well-defined crossing symmetry equations. These are the essential
ingredients for the numerical bootstrap program \cite{Ferrara:1973yt,Polyakov:1974gs,Rattazzi:2008pe,Rychkov:2009ij,Caracciolo:2009bx,ElShowk:2012ht, Kos:2014bka}, the analytic
results at large spin \cite{Fitzpatrick2013,Komargodski2013}, and more recently the analysis of causality
constraints \cite{Hartman2015,Hartman2016,Hofman2016}.

This work aims to use the exact same CFT structures to constrain non-conformal
quantum field theories. Our main vehicle for doing so is to place
the $D$-dimensional QFT in hyperbolic space, where the 
algebra of isometries $\mathfrak{so}(D,1)$ coincides with that of a CFT in $d= D-1$
dimensions. We will focus our investigations on \emph{boundary correlation
functions}; these can be defined as functional derivatives of the
bulk partition function with respect to the boundary conditions, or
alternatively by pushing the insertion points of bulk correlation
functions towards the conformal boundary. Such observables resemble
CFT correlation functions in almost all respects and the aforementioned
techniques can be applied straightforwardly. In this paper we will
investigate in particular the power of numerical bootstrap methods
to constrain these QFT observables.

The structure of QFTs in hyperbolic space forms an interesting subject
by itself, but for obvious reasons it would be more interesting if
the current setup would also allow us to determine flat-space observables
of the QFT. This leads us to consider the \emph{flat-space limit}
where we send the radius of curvature $R$ to infinity. In this limit
we would like to keep the masses of the bulk particles fixed, which
implies that the scaling dimensions $\Delta\sim mR$ of the dual boundary
operators will also diverge. We will discuss below how this brings
about interesting challenges for the numerical analysis.

Physically speaking we expect a close connection between the QFT S-matrix
and the flat-space limit of the boundary correlators in hyperbolic
space. The most concrete implementation of this idea comes through
the definition of a Mellin space transform of CFT correlators. As
explained in more detail below, there exists significant evidence
that the correct flat-space S-matrix can be reproduced from a simple
scaling limit of the Mellin transform of a boundary correlator.\footnote{Our prescription deviates slightly from earlier results, since the
external particles will be massive rather than massless in the flat-space
limit.} At the level of individual diagrams this procedure simply ``removes
the circle'' from a Witten diagram and transforms it into an ordinary
Feynman diagram, with external legs amputated as per the LSZ prescription.
We however expect the procedure to make sense more generally.  We also provide an alternative connection by expressing the flat-space phase shift directly in terms of (a limit of) the spectrum and OPE coefficients in the boundary correlation functions. This formula works only for physical values of the Mandelstam variables but has the significant advantage of making unitarity manifest.

This paper splits into two main parts. In the first, which comprises sections \ref{sec:qftinhypspace} and \ref{sec:flatspacelimit}, we will discuss
the physics of boundary correlators and the flat-space limit. We discuss general properties,
their differences and similarities to ordinary CFT correlation functions,
and our expectations for the flat-space limit and the connection to
the S-matrix. The second part consists of section \ref{sec:numericalresults}, where we apply numerical bootstrap techniques
to boundary correlators to investigate what information we can gather
both about QFTs in hyperbolic space and how to extrapolate to the
flat-space limit. In this initial exploration we have focused on two-dimensional QFTs in order to simplify the numerical analysis.

As explained further below, we have obtained very encouraging results. In particular we show that our construction allows for the extraction of \emph{upper bounds} on the residues of poles in a 2-to-2 elastic scattering amplitude of massive particles, which must be obeyed by any unitary two-dimensional QFT. We consider it highly nontrivial that such a result for \emph{massive} QFTs follows from an analysis of \emph{conformal} crossing symmetry equations. 

Our encouraging results led us to scrutinize the structure of the S-matrix for two-dimensional QFTs. As explained in our companion paper \cite{paper2}, it is in fact possible to \emph{directly} constrain the residues of poles in 2-to-2 elastic amplitudes, using only the assumptions of analyticity, crossing symmetry and unitarity and without resorting to a hyperbolic space construction. In sections \ref{sec:numericalresults} and \ref{sec:conclusions} we will discuss the excellent agreement between these two approaches and the ways in which they complement each other.

\section{QFT in hyperbolic space}
\label{sec:qftinhypspace}
The study of QFT in hyperbolic space is an old idea \cite{Callan1990}. In this section we review the salient features of this construction, with a focus on the definition of boundary operators and their correlation functions. In the next section we will discuss how these correlation functions will morph into an S-matrix in the flat-space limit.

\begin{figure}[t]
\includegraphics[width=16cm]{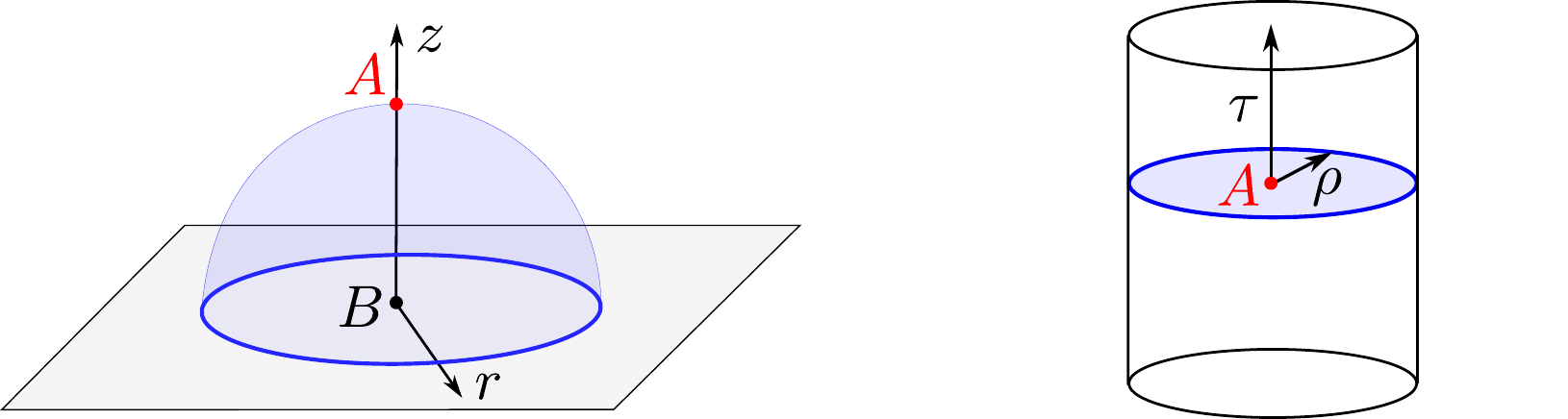}
\caption{Hyperbolic space in Poincar\'e coordinates (left) and global coordinates
(right). Surfaces of constant $\tau$ correspond to hemispheres of
radius $\sqrt{z^{2}+r^{2}}=e^{\tau}$ in the right picture. The boundary
point $B$ corresponds to $\tau=-\infty$ in global coordinates. \label{fig:Hyperbolic-space}}
\end{figure}

\textbf{The Box.} Hyperbolic space (also known as Anti-de Sitter space) is famed for introducing an IR cutoff while keeping the same number of isometries as in flat space. It can for example be described by the metric 
\[
ds^{2}=R^{2}\frac{dz^{2}+dr^{2}+r^{2}d\Omega_{d-1}^{2}}{z^{2}}\,.
\]
Here $R$ is the radius of curvature, $r$ is a radial coordinate for $\mathbb{R}^{d}$, and the coordinate $z > 0$.
These coordinates are useful because they give rise to a flat conformal
boundary at $z=0$, where the isometry group $SO(d+1,1)$ acts as
the conformal group on $\mathbb{R}^{d}$. Defining $z=e^{\tau}\cos\rho$
and $r=e^{\tau}\sin\rho$, we obtain AdS in global coordinates 
\begin{equation}
ds^{2}=R^{2}\frac{d\tau^{2}+d\rho^{2}+\sin^{2}\rho\,d\Omega_{d-1}^{2}}{\cos^{2}\rho}\,,\label{eq:globalAdS}
\end{equation}
where $\tau\in\mathbb{R}$ and $0<\rho<\frac{\pi}{2}$. 
These two coordinate systems are depicted in figure \ref{fig:Hyperbolic-space}.

\textbf{Boundary Operator/Bulk State Correspondence.} Surfaces of constant global time $\tau$ correspond to
hemispheres centered around the boundary point $z=r=0$ which shrink
to the boundary point $B$ in figure \ref{fig:Hyperbolic-space} when $\tau\to-\infty$. This picture leads to
a one-to-one map between states associated to surfaces of constant
global time $\tau$ and boundary operators inserted at $z=r=0$. 
On the one hand, the insertion of a boundary operator at $z=r=0$ prepares a state in the surface $\tau=0$. On the other hand, a state can be propagated backwards in time towards $\tau \to -\infty$ where it can be seen as a local operator inserted at the boundary point $B$.
We shall work in an eigenbasis of the Hamiltonian $H$ that generates
global time translations or, equivalently, dilatations around the
boundary point $B$. The states can be organized into representations
of the conformal group, which are labeled by the scaling dimension
$\Delta$ and the $SO(d)$ irreducible representation of the primary
state. For example, for a scalar particle of mass $m$ at rest in the center of AdS we have the familiar relation $\Delta (\Delta - d) = m^2 R^2$. 

\textbf{Bulk/Boundary Expansion.}
The boundary operators can be defined by pushing local bulk operators
towards the conformal boundary. More precisely, we can write a local
bulk operator $\phi_{i}$ as an infinite sum of boundary operators\footnote{We focus on scalar operators for simplicity.},
\[
\phi_{i}(z,x)=\sum_{k}a_{ik}\,z^{\Delta_{k}}\left[\mathcal{O}_{k}(x)+descendants\right]\,,
\label{AdSOBE}
\]
where $x\in\mathbb{R}^{d}$ is a cartesian coordinate on the flat
conformal boundary and we organized the sum into contributions from
the primary operators $\mathcal{O}_{k}$ and its descendants. It is easy to check that the action of the Killing vectors of hyperbolic space on the field $\phi$ induces the usual action of conformal generators on the primary operators $\mathcal O_k$.
The
(bulk state)-(boundary operator) map implies that this expansion has
a finite radius of convergence inside correlation functions. 

\textbf{Boundary Operator Product Expansion.} 
The same state-operator map leads to a convergent Operator Product Expansion (OPE) of the boundary operators
\begin{equation}
\mathcal{O}_{i}(x)\mathcal{O}_{j}(0)=\sum_{k}\lambda_{ijk}\,
\left|x \right|^{\Delta_{k}-\Delta_{i}-\Delta_{j}}\left[\mathcal{O}_{k}(0)+descendants\right]\,.\label{eq:OPE}
\end{equation}
%\begin{equation}
%\mathcal{O}_{i}(x)\mathcal{O}_{j}(0)=\sum_{k}C_{ijk}\,\left(x^{2}\right)^{\frac{\Delta_{k}-\Delta_{i}-\Delta_{j}}{2}}\left[\mathcal{O}_{k}(0)+descendants\right]\,.\label{eq:OPE}
%\end{equation}

\textbf{Conformal Theory.} The conclusion from the above discussion is therefore that any $d+1$ dimensional QFT in AdS$_{d+1}$ can be used to \textit{define} a set
of correlation functions that behave like correlators of a $d$ dimensional conformal theory (CT). We use this nomenclature to highlight that the boundary correlation functions of the $\mathcal{O}_{i}$'s do not define a conventional full-fledged conformal \textit{field} theory (CFT) simply due to the absence of operators like a stress tensor or currents for global symmetries in their OPE. (We discuss what happens to the bulk QFT stress tensor in appendix \ref{subapp:stresstensor}.) In any instance, the axioms of a conformal theory, most notably unitarity and the existence of a convergent OPE are all one needs to make use of conformal bootstrap techniques. 

Our setup differs from the standard AdS/CFT correspondence. There, the existence of a boundary stress tensor is well-known to correspond to the dynamical bulk metric. In this
paper we instead restrict ourselves to the study of QFT in a fixed AdS background geometry. 
{It might also be interesting to think about such QFTs as the limit
of bulk graviational theories where the Planck length was sent to
zero. This means that the set of boundary correlators we are studying
can be thought of as a sector of the dual CFT$_{d}$ in the limit
of infinite central charge. See for instance \cite{Aharony:2015zea} for a recent implementation of this idea.} 

\section{The flat space limit}\label{sec:flatspacelimit}
In hyperbolic space the radius of curvature $R$ acts as a finite-volume regulator, and for very large values of $R$ we naturally expect to recover the physics of infinite-volume flat space. In this section we discuss this limit in more detail. We will demonstrate that it translates into particular scaling limits for the conformal theory described in the previous section. This will lead us to formulate a precise dictionary between physical flat-space observables and CT data which we will bootstrap in the following section. 

The first element in this dictionary involves the masses in the flat space QFT and the dimensions of the CT. As discussed above, a scalar particle of mass $m_i$ in AdS can for example be created by a boundary operator of dimension given by $\Delta_i(\Delta_i-d)=m_i^2 R^2$. Therefore, by changing the AdS radius we smoothly vary the conformal dimensions of the CT. In this way we obtain a one-parameter family of CTs. We are interested in the limit where the the Compton wave length of the particle is much smaller than the AdS radius so that the particle perceives its surrounding as flat space. So we are interested in taking $m_i R \to \infty$ so that all dimensions of the CT should be taken to infinity with their ratios held constant. In this way we obtain the following simple relation between the dimensions of the operators of the CT and the masses of the particles (measured in units of the lightest particle)
\[
\boxed{\frac{m_{i}}{m_{1}}=\lim_{\Delta_{i}\to\infty}\frac{\Delta_{i}}{\Delta_{1}}.} \la{masses}
\]
Notice that it suffices to consider \emph{primary} boundary operators: these correspond to particles at rest whereas descendant states become boosted particles in the flat-space limit. In appendix \ref{appendix:rgflow} we discuss this limit in more detail, including the case where the QFT flows to a non-trivial IR fixed point. This discussion highlights rather sharply the distinction between a CFT -- where we have at least an operator (the stress tensor) with small anomalous dimensions -- and the CT's  under consideration -- where all operators acquire a parametrically large dimension -- alluded to at the end of the last section.

The second element of the dictionary relates flat space scattering amplitudes and correlation functions of the conformal theory. Here we propose two different relations for this dictionary, each with its own advantages and limitations.  

The first is most easily stated if we work in the Mellin representation \cite{Mack2009,Mack2009a}, whose definition is recalled below. The claim is that the $n$-particle flat scattering space amplitude can be directly extracted from the connected Mellin amplitude $M(\gamma_{ij})$ through the limit
\begin{equation}
\boxed{
(m_{1})^{a}\,T(k_{i})=\lim_{\Delta_{i}\to\infty}\frac{(\Delta_{1})^{a}}{\mathcal{N}}M\left(\gamma_{ij}= \frac{\Delta_{i}\Delta_{j}}{\Delta_{1}+\dots+\Delta_{n}}\left(1+\frac{k_{i}\cdot k_{j}}{m_{i}m_{j}}\right) \right)\label{eq:FSLgeneralformula}
}
\end{equation}
where $a=n(d-1)/2-d-1$ renders the expression dimensionless and where the normalization factor is given by a combination of gamma functions, 
\[
\mathcal{N}=\frac{1}{2}\pi^{\frac{d}{2}}\Gamma\left(\frac{\sum\Delta_{i}-d}{2}\right)\prod_{i=1}^{n}\frac{\sqrt{\mathcal{C}_{\Delta_{i}}}}{\Gamma(\Delta_{i})}\,, \qquad \mathcal{C}_{\Delta}\equiv\frac{\Gamma(\Delta)}{2\pi^{\frac{d}{2}}\Gamma\left(\Delta-\frac{d}{2}+1\right)}\,.
\]
A similar flat space limit formula
appeared before for the case of external massless particles \cite{Penedones:2010ue,Fitzpatrick:2011hu}.
It would be interesting to understand better the relation between
these two formulas. In particular, the flat space limit formula for
external massless particles involves an integral which is not present
in (\ref{eq:FSLgeneralformula}). 
We discuss further this relation, its derivation and its implications in subsection \ref{MellinSub}. 

We also found another relation between flat space scattering and the CT data. This second relation yields an expression for the spin $l$ phase shift $\delta_l(s)$ for a 2-to-2 S-matrix element describing the scattering of a particle of mass $m_1$ against a particle of mass $m_2$. The relation is even more direct than the previous one but only holds for physical values of the total energy in the center of mass frame~$\sqrt{s}=\sqrt{m_1^2+k^2}+\sqrt{m_2^2+k^2}$, that is for $\sqrt{s}>m_1+m_2$. It reads  
\[
\boxed{
e^{2i\delta_{l}(s)}=\lim_{\Delta_i \to \infty} 
\left.
\sum_{|\Delta-E|<\delta E}
\left[w(\Delta) \lambda_{\Delta,l}\right]^{2}e^{-i\pi(\Delta-\Delta_{1}-\Delta_{2}-l)}
 \Big/
  \sum_{|\Delta-E|<\delta E}\left[w(\Delta)\lambda_{\Delta,l}\right]^{2}
   \right.
\label{eq:phaseshiftformula1}
}
\] 
where $E/\Delta_1=\sqrt{s}/m_1$ is the center of mass energy measured in units of the lightest particle and $\lambda_{\Delta,l}$ are the OPE coefficients arising in the OPE of $\mathcal{O}_1$ and $\mathcal{O}_2$. (The weight $w$ is a simple function of the CT spectra discussed in detail below and the bin size $1\ll \delta E \ll E$.)
We discuss further this relation, its derivation and its implications in subsection \ref{secPhase}. 

Let us already anticipate that our derivations of these relations contain some heuristic elements and it would certainly be interesting to try to render them more rigorous. We  also did not rigorously establish the equivalence between these two formulas from a CT perspective, although we show in appendix \ref{sec:Mellintophaseshift} that formulas (\ref{eq:phaseshiftformula1}) and (\ref{eq:FSLgeneralformula}) give rise to the same imaginary part of the flat space scattering amplitude.

In section \ref{sec:numericalresults} we are going to analyze the large dimensions conformal theories from a bootstrap lens thus constraining the space of flat space massive quantum field theories. In practice we will use (\ref{masses}) and a particular restriction of  (\ref{eq:FSLgeneralformula}) to the three particle amplitude where this formula simplifies dramatically, see e.g. (\ref{eq:CtoG}) below. The reader curious about the bootstrap details might prefer to take the flat space formulae on faith on a first reading and jump directly to section \ref{sec:numericalresults}.

\subsection{Mellin Approach} \la{MellinSub}

The Mellin representation is a very useful Fourier transform of the four point correlation function with respect to the logarithm of the conformal cross-ratios.
We recall that Mellin amplitudes $M(\gamma_{ij})$ are defined \cite{Mack2009,Mack2009a} by expressing an $n$-point conformal correlation function as the integral
\be
\label{eq:Mellindefn}
\left\langle \mathcal{O}_{1}(x_{1})\dots\mathcal{O}_{n}(x_{n})\right\rangle =\int[d\gamma]M(\gamma_{ij})\prod_{1\le i<j\le n}\frac{\Gamma(\gamma_{ij})}{\left(x_i - x_j\right)^{2 \gamma_{ij}}}\,,
\ee
Here the Mellin variables $\gamma_{ij}$ obey the constraints
\begin{equation}
\gamma_{ij}=\gamma_{ji}\ ,\qquad\gamma_{ii}=-\Delta_{i}\ ,\qquad\sum_{i=1}^{n}\gamma_{ij}=0\ .
\end{equation}
These constraints can be solved in terms of $n(n-3)/2$ independent variables, which in \eqref{eq:Mellindefn} are integrated along a contour parallel to the imaginary axis as indicated by the symbol $[d\gamma]$. For theories in AdS space the Mellin amplitudes are particularly convenient \cite{Penedones:2010ue, Fitzpatrick:2011ia, Paulos:2011ie} and exhibit remarkable similarities with scattering amplitudes in flat space. This makes the Mellin amplitude a natural ingredient in our flat space relation \eqref{eq:FSLgeneralformula}.

In appendix \ref{sec:apendix:TfromM} we discuss several checks of equation \eqref{eq:FSLgeneralformula}. In particular, in section \ref{derivationPer} we verified equation \eqref{eq:FSLgeneralformula} for an \textit{arbitrary} contact term interaction using contact Witten diagrams. In principle this constitutes a derivation of (\ref{eq:FSLgeneralformula}). After all, we are dealing with massive particles so we can imagine integrating them out and generating in this way a plethora of effective contact term interactions. Since (\ref{eq:FSLgeneralformula}) holds for each of them it should hold for the sum over all possible interactions. Of course, things would be more subtle if we were dealing with massless particles. As a further cross-check we also verified (\ref{eq:FSLgeneralformula}) for a single scalar exchange in section \ref{sub:Scalar-exchange} and for a scalar loop diagram in section \ref{Two-particleSec}.

We can also adopt a slightly different point of view and take \eqref{eq:FSLgeneralformula} as a \textit{definition} of the bulk scattering amplitude in terms of the boundary correlator. This has some interesting conceptual consequences. 
The OPE implies a very simple analytic structure of Mellin amplitudes:
they are meromorphic functions with the position of the simple poles
fixed by the scaling dimension of the operators that appear in the
OPEs of the external operators. Moreover, the residues factorize into
(sums of) products of lower point Mellin amplitudes. As explained in detail in appendix \ref{sec:apendix:TfromM},
these analytic and factorization properties of the Mellin
amplitudes \textit{imply}, via formula (\ref{eq:FSLgeneralformula}), the expected
analytic and factorization properties of scattering amplitudes. One
can thus view this as a first principle \textit{derivation} of the S-matrix analyticity and factorization
axioms. 

On the other hand, unitarity of the S-matrix defined by (\ref{eq:FSLgeneralformula}) is not obvious.
This should follow automatically from unitarity of the boundary correlators (scaling
dimensions above conformal unitary bounds and real OPE coefficients)
however it is not clear what this implies for the Mellin amplitude. 
Fortunately, our second relation (\ref{eq:phaseshiftformula}) does render unitarity manifest and by relating the two formulas (see appendix  \ref{sec:Mellintophaseshift}) we explain unitarity of (\ref{eq:FSLgeneralformula}). 

For the remainder of the paper it is instructive to consider in detail the case of a three point function of scalar operators. In that case there is no independent Mellin variable, so (\ref{eq:FSLgeneralformula}) simplifies dramatically into a relation between the physical three-point couplings in flat space (measured in units of the lowest mass $m_1$) and the OPE coefficients of the boundary conformal theory,
\be
\boxed{
% (m_1)^{\frac{d - 5}2}  
 g_{123}  = \lim_{\Delta_i \to \infty}   \lambda_{123} \, \times  \frac{2 (\Delta_1)^{\frac{d - 5}2}}{\pi^{\frac{d}{2}}\Gamma(\frac{1}{2} \sum_{i=1}^3 \Delta_i - \frac{d}{2})} \prod_{i=1}^3 \frac{\Gamma(\Delta_i)}{\Gamma(\frac{1}{2} \sum_{i=1}^3 \Delta_i - \Delta_i)\,\sqrt{\mathcal C_{\Delta_i}}}\,.  \label{eq:CtoG}
}
\ee

\begin{figure}[t]
\begin{center}
\includegraphics{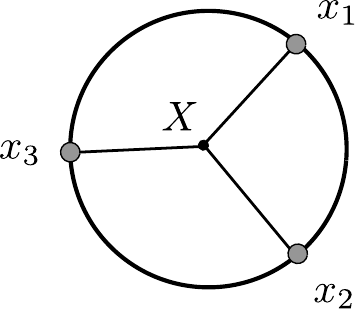}
\caption{\label{fig:3ptwittendiagram}Three-point Witten diagram.}
\end{center}
\end{figure}

We can in fact re-derive this relation, independently of any Mellin transform, by considering the case of three weakly coupled scalar fields $\phi_i$, $1 \leqslant i \leqslant 3$, with
a cubic vertex $\hat g_{123}\, \phi_1 \phi_2 \phi_3$ in AdS$_{d+1}$. Notice that the coupling $\hat g_{123}$ is dimensionful; we measure it in units of the mass of the lightest particle so our dimensionless coupling is $g_{123}= \hat g_{123}/ m_1^{(5-d)/2} $.  The scaling
dimension $\Delta_i$ of the boundary operators is related to the mass
$m_i$ of the scalar field $\phi_i$ via $m_i^{2}R^{2}=\Delta_i(\Delta_i-d)$.
The tree level boundary three-point function is given by the
Witten diagram shown in figure \ref{fig:3ptwittendiagram}. 
This gives 
\begin{equation}
\left\langle \mathcal{O}_1(x_{1})\mathcal{O}_2(x_{2})\mathcal{O}_3(x_{3})\right\rangle =g_{123} (m_1 R)^{\frac{5-d}{2}}\int_{0}^{\infty}\frac{dz}{z^{d+1}}\int d^{d}x\prod_{i=1}^{3}\frac{
\sqrt{\mathcal{C}_{\Delta_i}}z^{\Delta_i}}{\left[z^{2}+(x-x_{i})^{2}\right]^{\Delta_i}}\,,\label{eq:cubicWittendiagram}
\end{equation}
where, as above, 
\[
\mathcal{C}_{\Delta}=\frac{\Gamma(\Delta)}{2\pi^{\frac{d}{2}}\Gamma\left(\Delta-\frac{d}{2}+1\right)}
\]
arises from normalizing the boundary operators to have unit two point
function. On the other hand, the boundary three point function is
fixed by conformal symmetry up to an overall constant, 
\[
\left\langle \mathcal{O}_1(x_{1})\mathcal{O}_2(x_{2})\mathcal{O}_3(x_{3})\right\rangle 
=\frac{ \lambda_{123} }
{x_{12}^{\Delta_{12,3}} x_{13}^{\Delta_{13,2}} x_{23}^{\Delta_{23,1}}}\,, \label{c123Ex}
\]
where $x_{ij} = |x_{i}-x_{j}|$ and $\Delta_{ij,k} = \Delta_i + \Delta_j - \Delta_k$. With our normalizations, this
constant is just the OPE coefficient appearing in (\ref{eq:OPE}).
The integral in (\ref{eq:cubicWittendiagram}) was computed already in \cite{Freedman:1998tz}. By equating the result to (\ref{c123Ex}) and taking the flat space limit corresponding to large external dimensions, we precisely recover~(\ref{eq:CtoG}).

Finally let us quote here a particular example of the above relation which will be used extensively in the bootstrap of section \ref{sec:numericalresults}. Consider the coupling between two particles of mass $m_1$ and third particle of mass $m_2 =\alpha\, m_1$. (So that in the CFT we have a correlator between two operators of large dimension $\Delta_1$ and a third operator of dimension $\Delta_2= \alpha \Delta_1$, also large.) In this case (\ref{eq:CtoG}) can be simplified into 
\be
g_{112} = 2^{-\frac{d}{2}}  \pi^{\frac{d-2}{4}}  (2-\alpha)^{1/2} \alpha ^{1-\frac{d}{4}} (\alpha +2)^{\frac{d+1}{2}} \lim_{\Delta_i\to\infty} \Delta_1^{\frac{d-2}{4}} \left( 2^{\alpha +2} (2-\alpha )^{\frac{\alpha -2}{2}} (2 + \alpha)^{\frac{-\alpha -2}{2}}\right)^{\Delta_1} \lambda_{112} 
\label{eq:FSLcubiccoupling}
\ee
The term in parentheses is positive and greater than one, so 
we see that a finite cubic coupling $g_{112}$ corresponds to an OPE
coefficient $\lambda_{112}$ that decays \emph{exponentially}
with $\Delta_1$. This scaling is generic (and unrelated to this particular example) and agrees with the findings of \cite{Bargheer:2013faa,Minahan:2014usa}. As explained there, it has a simple physical explanation. Basically, since the dimensions are very large the propagation from the boundary to the bulk is governed by a semi-classical approximation and leads to an exponential weight $e^{-m_1 \mathcal{L}_1-m_2 \mathcal{L}_2-m_3 \mathcal{L}_3}$ where $\mathcal{L}_i$ are the (renormalized) length of geodesics connecting the boundary points to an interaction point in the bulk (whose location maximizes this weight). To measure the flat space coupling felt by the particles when they reach this interaction point we should thus strip out this exponential factor as in (\ref{eq:FSLcubiccoupling}).

This physical picture -- with particles propagating in the bulk until they meet in a small region where they effectively interact as in flat space -- also explains why any conformal bootstrap numerics should be quite challenging. Consider a four point correlation function of, say, identical boundary operators $\mathcal{O}$. Its leading contribution will be given by the disconnected contribution where particles fly from one boundary point to another without any interaction. The interesting part of the result, on the other hand, is the connected contribution which is exponentially smaller.  To extract a flat space S-matrix we need therefore to subtract out the huge disconnected background from the connected contribution which in turn is exponentially small. We should then strip out the exponentially small propagation weights to finally get an order $1$ amplitude in flat space. On top of all this we must then extrapolate the results of the numerics towards the limit when all dimensions are scaled to infinity so that the AdS box becomes effectively flat space! In practice this translates into the necessity of keeping hundreds of digits of precision in any bootstrap numerics to obtain just a few digits of precision for the flat-space result. It is the price to pay for such a cool scattering Gedankenexperiment.

\subsection{Phase shift} \la{secPhase}

In the case of 2 to 2 scattering there is an alternative way to obtain the scattering amplitude. The idea is to consider the phase shift $\delta_l(s)$ given by
\be
e^{2i\delta_l(s)} = 
\ _{out}\langle s,l| s,l \rangle_{in} =
 \ _{in}\langle s,l|\hat{S}| s,l \rangle_{in} 
\ee
where $\hat{S}$ is the S-matrix and 
\be
|s,l\rangle_{in} \propto \int_{S^{d-1}} d\vec{n}\, P_l(\vec{n} \cdot \vec{n}_0) \, |\vec{k}_1= k \vec{n}, \vec{k}_2=-k \vec{n} \rangle_{in}
\ee
is a two-particle eigenstate of angular momentum. Here, $\sqrt{s}=\sqrt{m_1^2+k^2}+\sqrt{m_2^2+k^2}$ is the total energy in the center of mass frame, $\vec{n}_0$ is an arbitrary unit vector defining a reference axis and   $P_l(\vec{n} \cdot \vec{n}_0)$ is the degree $l$ harmonic polynomial on the sphere $S^{d-1}$ at spatial infinity. In this language, unitarity is the simple statement
\[
\left|e^{2i\delta_{l}(s)}\right|\le1 \qquad\text{for}\qquad s\ge(m_1+m_2)^2\,.
\]

This construction has a simple analogue in the case of QFT in AdS.
Consider the following bulk state
\[
|\Psi_{l}\rangle=\int_0^1dx_1 dx_2 f(x_1,x_2) \left[\mathcal{O}_{1}(x_1)\mathcal{O}_{2}(-x_2)|0\rangle\right]_{primaries\ of\ spin\ l}
\label{statespinprojected}
\]
produced by the insertion of two boundary operators inside the unit sphere (see figure \ref{fig:Phaseshift}) and projected onto the space of primary operators of spin
$l$. In appendix \ref{sec:scatteringstates} we explain what is the appropriate weight $f(x_1,x_2)$ that can be used to produce scattering states in AdS. Given this weight function, we find that our state $|\Psi_{l}\rangle$ has a simple expansion in eigenstates of the cylinder
hamiltonian,\footnote{The state $|\Delta,l\rangle=\hat{n}^{\mu_1}\dots \hat{n}^{\mu_l}|\Delta,\{\mu_1, \dots, \mu_l\} \rangle$ is a particular component of the $SO(d)$ multiplet of primary states of dimension $\Delta$ and spin $l$.}
\[
|\Psi_{l}\rangle=\sum_{\Delta}w(\Delta)\lambda_{\Delta,l}|\Delta,l\rangle\,,
\label{stateexpandedineigenstates}
\]
where
\be
  w(\Delta)= \left[
\frac{4\Delta^2 (\Delta-\Delta_1-\Delta_2) }{
(\Delta^2-\Delta_{12}^2)(\Delta+\Delta_1+\Delta_2) }
\right]^{\frac{\Delta}{2}}
\frac{
\left(\frac{\Delta-\Delta_{12}}{\Delta+\Delta_{12}}\right)^{\frac{\Delta_{12}}{2}} }{
\left[ \Delta^2 -(\Delta_1+\Delta_2)^2 \right]^{\frac{\Delta_1+\Delta_2}{2}} }
\ ,
\ee
with $\Delta_{12}=\Delta_1-\Delta_2$,
 and $\lambda_{\Delta,l}$ are the OPE coefficients appearing in $\mathcal{O}_1 \times \mathcal{O}_2$. 

Let us now also project onto primaries with dimension $\Delta\in]E-\delta E,E+\delta E]$ for $E \gg \delta E \gg 1$ 
and normalize the state. We obtain:
\[
|\Psi_{l}(E)\rangle=\frac{1}{\sqrt{N_{l}(E)}}\sum_{|\Delta-E|<\delta E}w(\Delta)\lambda_{\Delta,l}|\Delta,l\rangle\,,
\qquad\qquad
N_{l}(E)=\sum_{|\Delta-E|<\delta E}\left[w(\Delta)\lambda_{\Delta,l}\right]^{2}\,.
\]
By construction this state has angular momentum $l$ and energy approximately $E$ in AdS. Moreover, it does not have center of mass motion due to the primary condition.

\begin{figure}
\centering
\includegraphics[width=14cm]{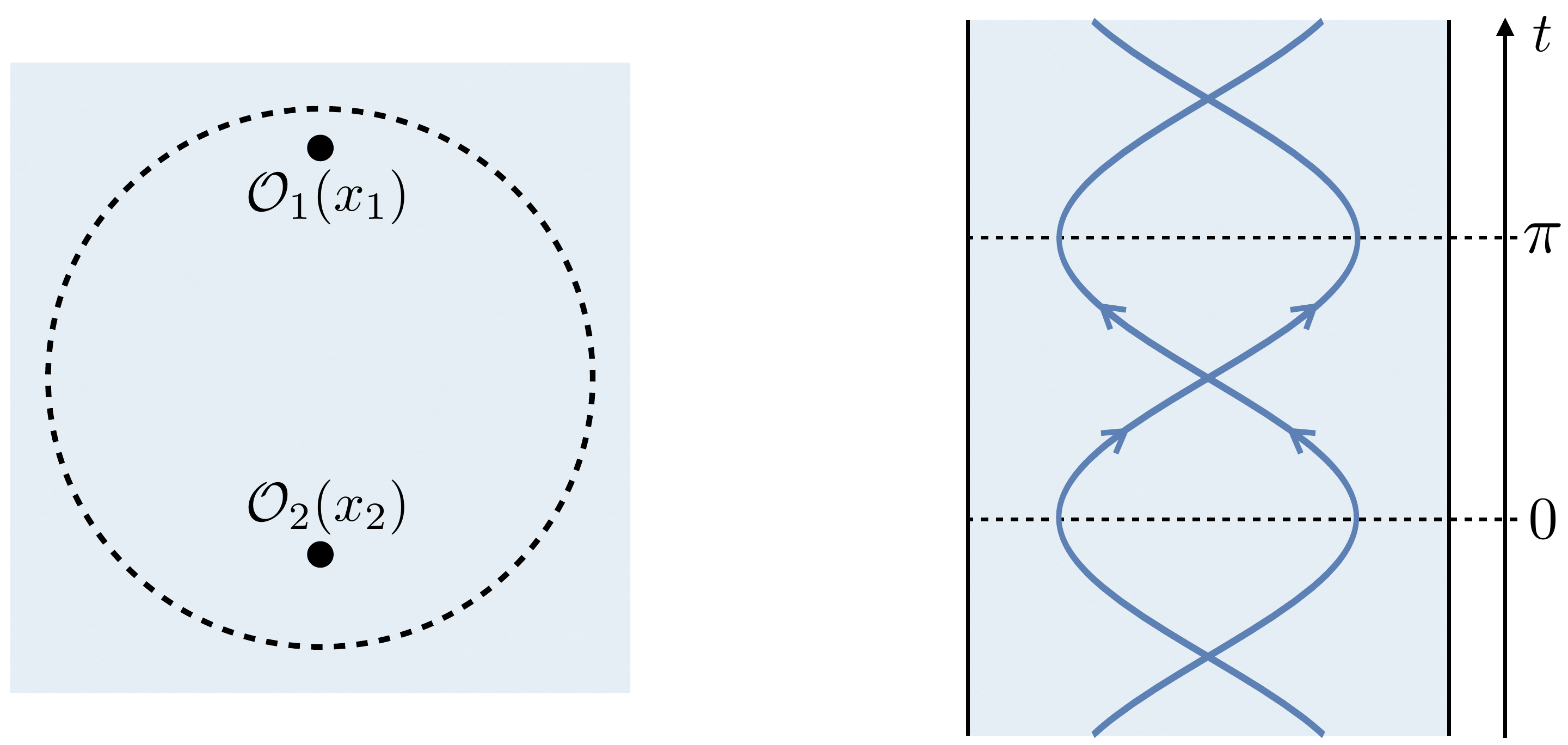}
\caption{On the left, we show the euclidean preparation of the two particle scattering state  by the insertion of $\mathcal{O}_1$ and $\mathcal{O}_2$ inside the unit sphere. Projecting onto primaries of spin $l$ and dimension $\Delta \in ]E-1,E+1]$ and taking the limit $x^2\to 1$ we obtain the state
$|\Psi_l(E)\rangle$.
On the right, we depict the Lorentzian evolution of this state starting from $t=0$. The blue lines indicate timelike geodesics that represent the classical evolution of two massive particles in AdS in the center of mass frame. The periodicity of these geodesics leads to a scattering event  for each time interval $\Delta t =\pi$.
 \label{fig:Phaseshift}}
\end{figure}

If the bulk theory is free then the correlation functions of the boundary operators $\mathcal{O}_{1}$ and $\mathcal{O}_{2}$ reduce to products of two-point functions. In this case, the OPE  $\mathcal{O}_1 \times \mathcal{O}_2$ will only include operators with dimension $\Delta=\Delta_{1}+\Delta_{2}+l+2n$
for $n=0,1,2,\dots$. Therefore, the  state described above is the state
\[
|\Psi_{l}(E)\rangle=\frac{1}{n_+ - n_-+1} \sum_{n=n_-}^{n_+} |\Delta_{1}+\Delta_{2}+l+2n,l\rangle\,,
\label{free2particle}
\]
with $n_{\pm}$ being the closest integer to $\frac{E\pm \delta E-\Delta_{1}-\Delta_{2}-l}{2}$. The state $|\Delta_{1}+\Delta_{2}+l+2n,l\rangle$ describes two non-interacting particles in AdS with relative angular momentum $l$ and radial quantum number $n$.

Timelike geodesics in AdS are periodic. This periodicity gives rise to one scattering event per global time interval $\Delta\tau=\pi$, as depicted in figure \ref{fig:Phaseshift}.
Therefore, we should define the scattering phase shift by
\[
\langle\Psi_{l}(E)|e^{-i\pi (H-H_0)}|\Psi_{l}(E)\rangle=\langle\Psi_{l}(E)|e^{-i\pi(H-\Delta_{1}-\Delta_{2}-l)}|\Psi_{l}(E)\rangle
\]
where $H_0$ is the free hamiltonian.
In the flat space limit, we find
\[
\boxed{
e^{2i\delta_{l}(s)}=\lim_{E\to \infty} \frac{1}{N_{l}(E)}\sum_{|\Delta-E|<\delta E}
\left[w(\Delta) \lambda_{\Delta,l}\right]^{2}e^{-i\pi(\Delta-\Delta_{1}-\Delta_{2}-l)}
\label{eq:phaseshiftformula}
}
\]
with the ratios $E/\Delta_1=\sqrt{s}/m_1$ and $E/\Delta_2=\sqrt{s}/m_2$ fixed, $\delta E \to \infty$ and $\delta E/ E \to 0$.
This gives a very direct relation between the CT data and the scattering data of the bulk theory in flat space. In addition, this formula makes unitarity manifest. In particular we see that absorption, \emph{i.e.} $\left|e^{2i\delta_{l}(s)}\right|<1$, corresponds to the existence of  several spin $l$ operators in the band $\Delta\in]E-\delta E,E+\delta E]$ with dimensions that do not differ by even integers. In this case, $|\Psi_{l}(E)\rangle$ does not come back to itself after evolving for the time interval $\pi$. In equation (\ref{eq:phaseshiftformula}), the phases $e^{-i\pi(\Delta-\Delta_{1}-\Delta_{2}-l)}$ will not be aligned and there will be absorption.

If the QFT  is weakly coupled, the two particle states (\ref{free2particle}) only get a small energy shift.
In the language of CT this corresponds to small anomalous dimensions $\gamma(n,l)$ for the boundary operators.
There can also be new small OPE coefficients $\tilde{\lambda}_{\Delta,l}$ that appear at leading order.
In this case, (\ref{eq:phaseshiftformula}) simplifies to the relation 
\[
2\delta_{l}(s)=-\pi\lim_{n\to \infty}\gamma(n,l)+
\lim_{E\to \infty} \frac{i}{N_{l}(E)}\sum_{|\Delta-E|<\delta E}
\left[w(\Delta)\tilde{\lambda}_{\Delta,l}\right]^{2}\left[1-e^{-i\pi(\Delta-\Delta_{1}-\Delta_{2}-l)}\right]
\,,
\]
with $\frac{\Delta_1+\Delta_2+2n}{\Delta_1}=\frac{\sqrt{s}}{m_1}$ and $\frac{\Delta_2}{\Delta_1}=\frac{m_2}{m_1}$ fixed.  
This is very similar to expressions that appeared previously \cite{Cornalba:2007zb, Heemskerk:2009pn, Fitzpatrick:2010zm}.
Notice that the second term gives a contribution localized at $\sqrt{s}= m_1\Delta /\Delta_1$ where $\Delta$ is the dimension of the new operators that appear in the OPE when we turn on a weak interaction.

This construction gives the phase shift in flat space directly from
a limit of the CT data. However, this only works for physical energy
$E>\Delta_{1}+\Delta_{2}+l$ and does not teach us about the analytic
structure of $\delta_{l}(s)$. 
In appendix \ref{sec:Mellintophaseshift}, we explain how formula \eqref{eq:phaseshiftformula} is related to the Mellin space formula \eqref{eq:FSLgeneralformula}.
Since unitarity is obvious in \eqref{eq:phaseshiftformula} this relation provides an argument for unitarity of our Mellin space formula  \eqref{eq:FSLgeneralformula}.
Moreover, in appendix \ref{sec:Mellintophaseshift} we argue that the spectral density $N_l(E)$ is universal in the flat space limit and, therefore, is the same as for free fields in AdS.

\section{Conformal Theory Bootstrap} 
\label{sec:numericalresults}
Let us pause to summarize what we have learned so far. Firstly, a quantum field theory in AdS has a natural set of observables, namely the boundary correlation functions, which have exactly the same structure as those of an ordinary conformal field theory (except that they do not feature a stress tensor). Secondly, we have argued that in a specific limit, corresponding roughly to sending the AdS radius to infinity, these observables transform into the flat-space S-matrix of the bulk QFT. We can therefore understand flat-space, non-conformal physics by studying the appropriate limit of the boundary correlators captured by the conformal theory.

Conformal correlation functions are subject to the well-known crossing symmetry equations. Fortunately for us, in recent years a growing body of work has shown that these equations can be mined very effectively to constrain the fundamental observables in unitary CFTs, \emph{i.e.}~the operator scaling dimensions and the OPE coefficients. Here we will focus on using numerical bootstrap methods to obtain results for massive unitary QFTs. Other possible analyses of the crossing symmetry equations, for example based on analytic methods, will be left to future work.

In this paper we focus on \emph{two-dimensional} QFTs. In that case the boundary correlators live in a \textit{one-dimensional} space, which brings about significant numerical simplifications: there are no spinning operators and four-point functions involve only a single cross-ratio. We expect to report results for higher-dimensional QFTs in the near future.

\subsection{Setup}
\begin{figure}[t]
\begin{center}
\includegraphics[width=5.5cm]{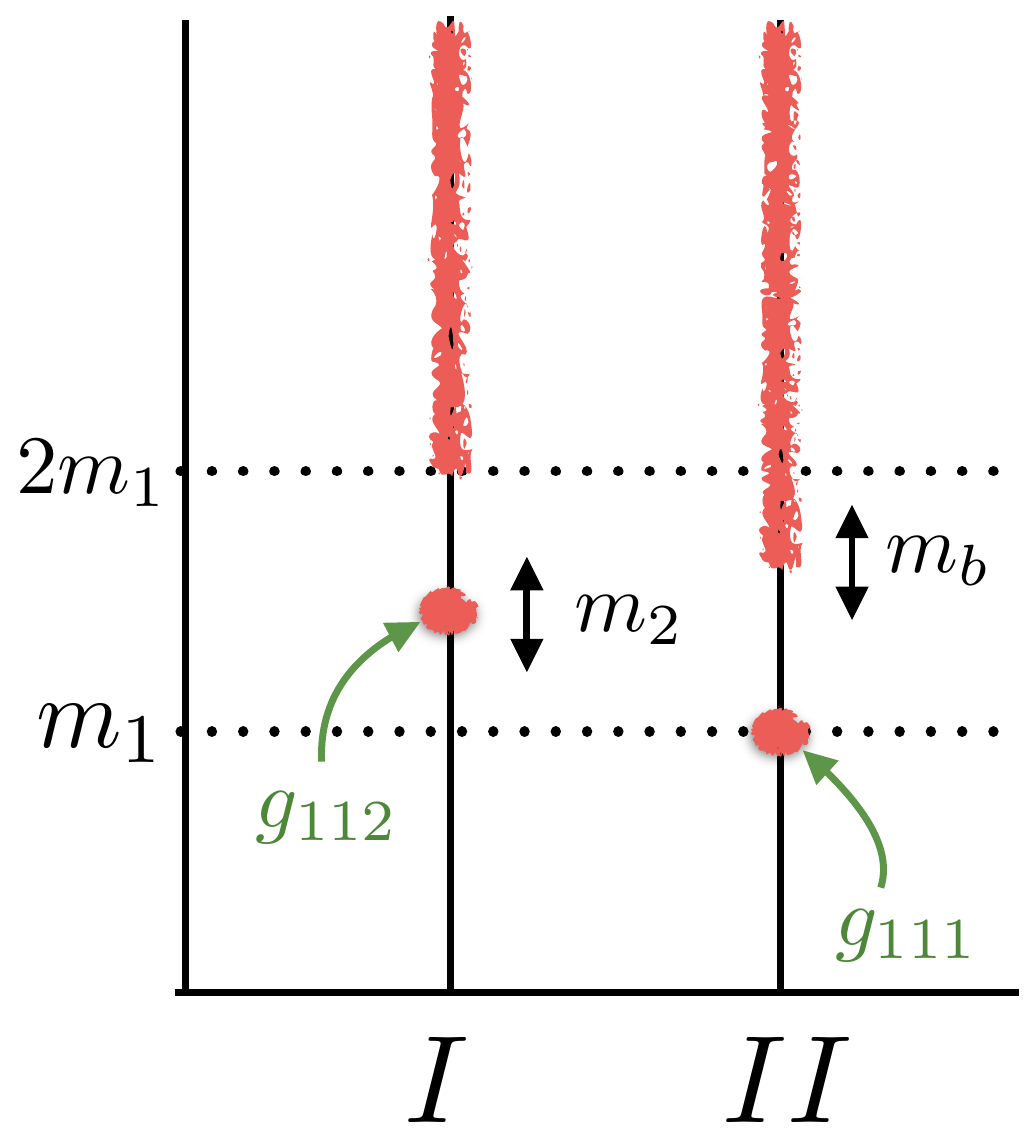}
\end{center}
\caption{\label{fig:scenarios}In scenario I we vary $m_2$ and find an upper bound on $g_{112}$. In scenario II we vary $m_b$ and find an upper bound on $g_{111}$. In both scenarios the mass of the scattered particle is $m_1$.}
\end{figure}

Let us consider a two-dimensional unitary QFT with a lightest stable massive scalar particle of mass $m_1$, and focus on the spectrum appearing in the elastic scattering of two such particles. In flat space this scattering event is described by the S-matrix element $S_{11 \to 11}$, but in AdS it is described instead by the CT four-point function
\be
\langle \cO_1(x_1) \ldots \cO_1(x_4)\rangle
\ee
of an operator $\cO_1$ with dimension such that $\Delta_1 (\Delta_1 -d) = m_1^2$. We will submit this four-point function to a numerical analysis in two following scenarios, displayed in figure \ref{fig:scenarios}.\footnote{In our companion paper \cite{paper2} we use a slightly different notation: $m_1$ here becomes $m$ there, and $m_2$ and $m_b$ here become $m_1$ there.}

\begin{itemize}
\item In scenario I we assume that the S-matrix has \emph{a single pole} corresponding to a particle with mass $m_2$ and then is analytic all the way up to the two-particle continuum at $2 m_1$. This scenario translates into a CT with an OPE of the form:
\be
\boxed{
\text{scenario I:} \qquad \cO_1 \times \cO_1 = 1 + \lambda_{112} \cO_2 + \ldots \text{(operators with $\Delta > 2 \Delta_1$)} \ldots
}
\ee
with $\Delta_i(\Delta_i - d) = m_i^2$. The squared OPE coefficient $\lambda_{112}^2$ corresponds via equation \eqref{eq:CtoG} to the residue at the pole, which we denote as $g_{112}^2$. We will be able to obtain an upper bound on this coefficient as a function of the dimensionless mass ratio $m_2/m_1$.\footnote{Notice that for $m_2 \neq m_1$ we assume in particular that the three-point coupling $g_{111} = 0$. In realistic theories this might be due to a symmetry, but we do not have to commit to any specific underlying mechanism.} The physical intuition behind this scenario is that the exchanged particle with mass $m_2$ mediates an attractive force between the particles of mass $m_1$ with a strength that is parametrized by $g_{112}^2$. If this interaction would be very strong then we would expect a bound state to form, which would manifest itself as an additional pole in the S-matrix and an operator of dimension $\Delta_2 < 2 \Delta_1$ in the CT. Since we assume that such a state is absent, we have the right to expect an upper bound on $g_{112}^2$.

\item In scenario II we assume instead that the S-matrix has a pole with residue $g_{111}^2$ that corresponds to a self-coupling of the scattered particle, and then no other poles up to a certain threshold which we will call $m_b$. In the CT language this becomes
\be
\boxed{
\text{scenario II:} \qquad \cO_1 \times \cO_1 = 1 + \lambda_{111} \cO_1 + \ldots \text{(operators with $\Delta > \Delta_b$)} \ldots
}
\ee
with the same translations to flat-space quantities as before. We will again obtain an upper bound on the residue $g_{111}^2$, now as a function of the dimensionless ratio $m_b/m_1$. In this case we can heuristically think of $m_b$ as the mass of a bound state of two $m_1$ particles. Since the binding strength is once more parametrized by the resiude $g_{111}^2$, we now not only expect to find an upper bound on $g_{111}^2$ but also that it will \emph{decrease} as we increase $m_b/m_1$. This intuition will be borne out below.
\end{itemize}
The attentive reader will have noticed that scenarios I and II coincide at the single point when $m_2 = m_b /2 = m_1$.

In order to obtain the desired upper bounds on the squared OPE coefficients $\lambda_{11 \ldots}^2$ we made use of the well-established numerical bootstrap algorithms \cite{Rattazzi:2008pe}. The basic idea is always to start with the conformal block decomposition of the four-point function, which in one dimension takes the form:
\be
\langle \cO_1(0) \cO_1(z) \cO_1(1) \cO_1(\infty)\rangle = \frac{1}{z^{2 \D_1}} \sum_k \lambda_{11k}^2 G_{\Delta_k}(z)\,,
\ee
with
\be
G_{\Delta_k}(z) \colonequals z^{\Delta_k} {}_2 F_1(\Delta_k,\Delta_k,2\Delta_k,z)\,,
\ee
and with $z = (x_{12} x_{34}) / (x_{13}x_{24})$ the only independent cross-ratio. Since all four operators are identical the four-point function obeys the crossing symmetry equation
\be
 \sum_k \lambda_{11k}^2 \left( \frac{1}{z^{2 \D_1}} G_{\Delta_k}(z) - (z \to 1-z) \right) = 0
\ee
Following standard procedures, we act on this equation with a linear functional $\alpha$. By linearity we obtain:
\be
\label{eq:alphaoncrossing}
 \sum_k \lambda_{11k}^2 \,\alpha \cdot \left[ \frac{1}{z^{2 \D_1}} G_{\Delta_k}(z) - (z \to 1-z) \right] = 0
\ee
Since the $\lambda_{11k}^2$ are positive, it is possible to find functionals which lead to impossibilities under certain assumptions for the structure constants and/or the spectrum. This in turn allows one to rule out such assumptions and thereby establish rigorous bounds.

For our numerical investigations we adopted the conventional form for $\alpha$, namely
\be
\alpha \cdot [f(z)] \colonequals \alpha_1 f'(1/2)+ \alpha_2 f''(1/2) + \dots+\alpha_N f^{(N)}(1/2)\,.
\ee 
The even derivatives vanish identically in \eqref{eq:alphaoncrossing} so the finitely many real numbers $\alpha_{2i - 1}$, $i \in {1,2,\ldots,\lfloor (N+1)/2 \rfloor}$ completely parametrize our functional. As $N$ increases the class of functionals we work with becomes more general and the bounds get better. Of course, searching for functionals for larger values of $N$ also requires greater computational resources. As indicated in the various plots below, our results were obtained with $N \lesssim 200$ for scenario I and with $N \lesssim 300$ scenario II.

The specific algorithm to constrain OPE coefficients was first introduced in~\cite{Caracciolo:2009bx}. Current state-of-the-art methods have been encoded in specialized software packages like {\tt JuliBootS}~\cite{Paulos2014a} and {\tt SDPB}~\cite{Simmons-Duffin2015}, both of which were used to obtain the results discussed below. For the high-precision results of scenario II we also made essential use of the `flow' method discussed recently in \cite{El-Showk:2016mxr}.\footnote{Full details of the numerical implementations are available from the authors upon request.} Notice that the flat-space limit dictates that our main interest is the behavior of the numerical bounds as $\Delta \to \infty$, which is very different from the usual searches where $\Delta$ is usually $O(1)$. For large $\Delta$ the numerical bootstrap analysis is unfortunately less efficient, as evidenced both by our numerical results and the $\Delta \gg N^2$ analysis in appendix \ref{appendix:largedelta}. We will therefore resort to an extrapolation procedure that we explain below.

\subsection{Results for scenario I}
\begin{figure}[t]
\begin{center}
\includegraphics[width=13cm]{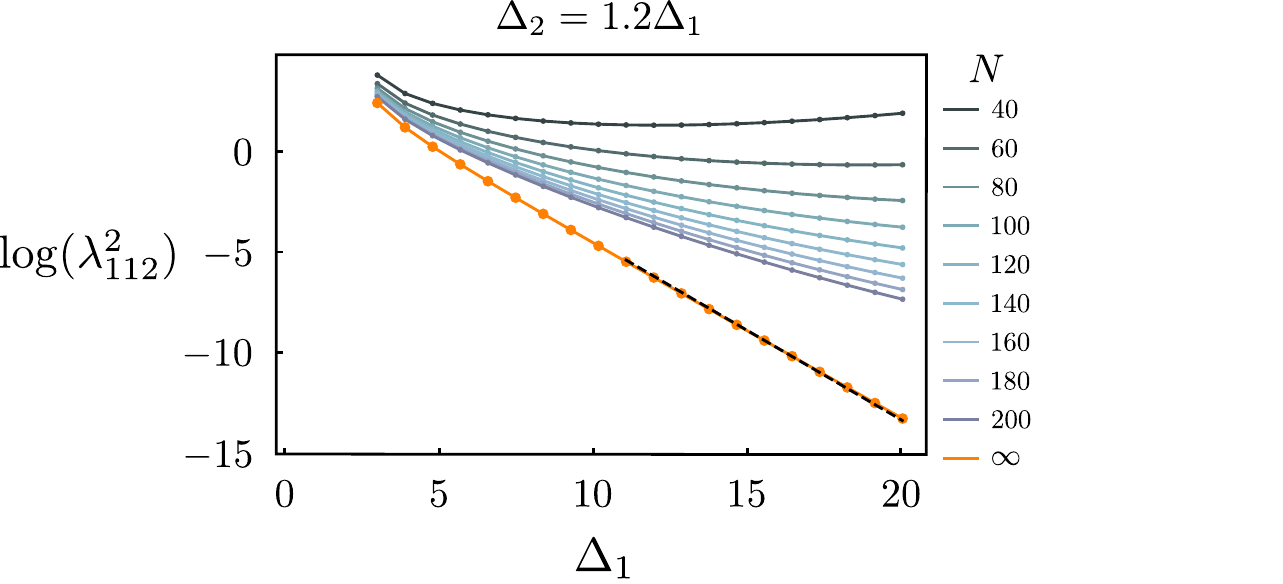}
\caption{\label{fig:scenarioIplot1}Numerical bounds for scenario I in the specific case with $\Delta_2 = 1.2 \Delta_1$. The orange line is the extrapolation of the numerical results to $N = \infty$ and for large $\Delta_1$ it accurately matches the expected flat-space slope as indicated by the dashed line.}
\end{center}
\end{figure}

\begin{figure}[t]
\begin{center}
\includegraphics[width=14cm]{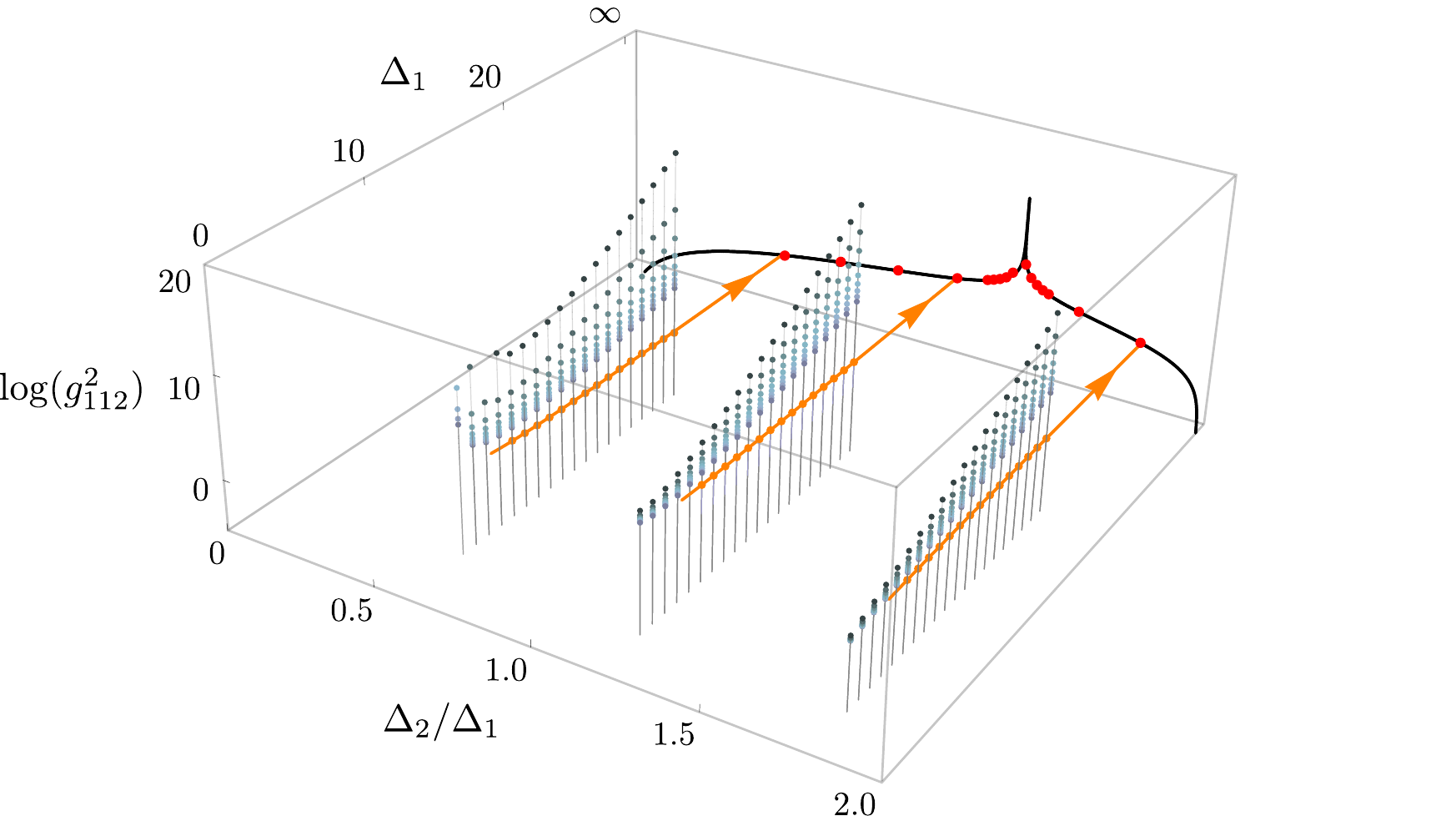}
\end{center}
\caption{\label{fig:extrapolations}Visualization of the double extrapolation procedure. Blue dots: some of our raw data points; each column of points corresponds to a series obtained with increasing $N$. Orange dots: extrapolation to $N = \infty$. Orange lines: fits and extrapolations to infinite $\Delta$. Similar extrapolations from data not shown lead to the series of red dots at the back surface. These constitute our main result and are shown independently in figure \ref{fig:scenarioIplot2}. Black line: exact result from the S-matrix bootstrap \cite{paper2} which tracks our numerical result.}
\end{figure}

\begin{figure}[t]
\begin{center}
\includegraphics[width=15cm]{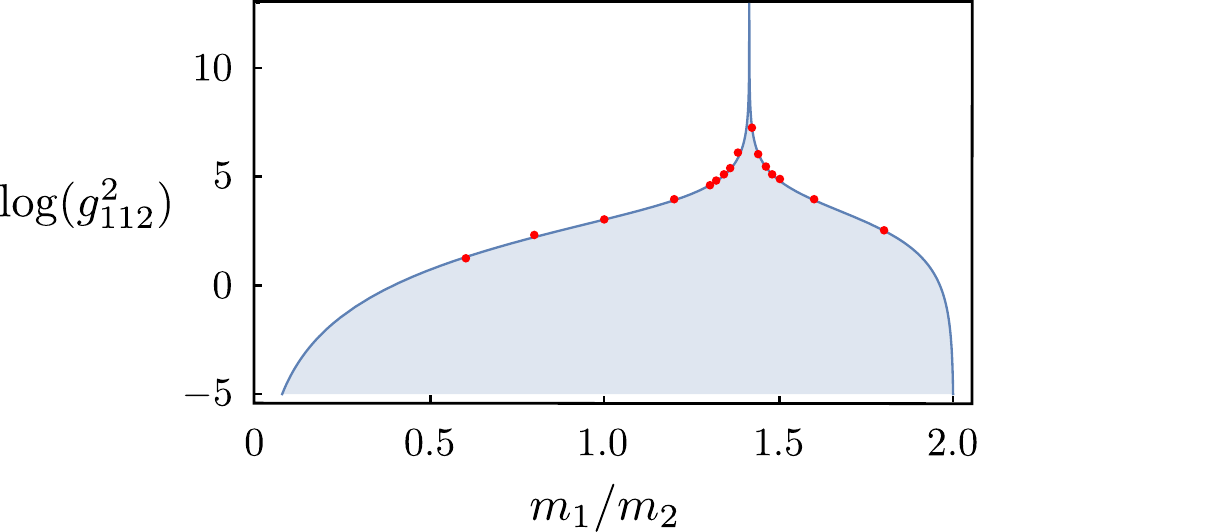}
\caption{\label{fig:scenarioIplot2}The double extrapolation to infinite $N$ and to the flat-space limit results in the plotted points, each one corresponding to a different ratio $\Delta_2/\Delta_1 \simeq m_2/m_1$. The blue curve is the exact result of the S-matrix bootstrap obtained in our companion paper \cite{paper2}.}
\end{center}
\end{figure}

We need to perform various extrapolations of our numerical results to obtain a physically relevant answer. We will therefore begin by explaining this procedure using figures \ref{fig:scenarioIplot1} and \ref{fig:extrapolations}; our final result is shown in figure \ref{fig:scenarioIplot2}.

Let us begin with the blue data in figure \ref{fig:scenarioIplot1}, which are our `bare' results for the specific representative case with $\Delta_2 = 1.2 \Delta_1$. Different lines correspond to different computational complexity as parametrized by $N$. It is clear that our bounds still heavily depend on $N$, especially for large $\Delta_1$. In order to get physically interesting results we therefore extrapolate the bounds to $N = \infty$, using a degree eight polynomial in $N^{-1}$. The result of such an extrapolation is represented by the larger orange points; this is our prediction for the upper bound that we would have obtained with infinite computational resources.\footnote{As cross-checks on the extrapolation procedure, we have checked that extrapolation using smaller values of $N$ can reproduce the results of higher values, and also that the final answer does not sensitively depend on the degree of extrapolation or which exact values of $N$ one includes.}

The next step is to translate the upper bound on $\lambda_{112}^2$ to an upper bound on the flat-space coupling $g_{112}^2$ using the results of the previous sections. For the plotted data we can use equation \eqref{eq:FSLcubiccoupling} with $\alpha = 1.2 = 6/5$. To leading order this results in
\be
\log(g_{112}^2) = \log(\lambda_{112}^2) + 2 \Delta_1\log(25/16) + O(\log(\Delta_1))\,.
\ee
The dashed line in figure \ref{fig:scenarioIplot1} is a least-squares fit to the orange data of a straight line with slope $ -2 \Delta_1 \log(25/16)$. The good fit is our first indication of success and puts us in an excellent position to extract an upper bound on $g_{112}^2$ for the flat-space S-matrix. We note that the results for other values of the ratio $\Delta_2 / \Delta_1$ show very similar behavior.

The true flat-space bound is then obtained by a further extrapolation to large $\Delta_1$ as visualized in figure \ref{fig:extrapolations}. Although we have obtained bare data for 19 different ratios of $\Delta_2/\Delta_1$, for clarity of presentation we have chosen to show only the bare data for $\Delta_2 /\Delta_1$ equal to $3/5$, $6/5$ and $9/5$. Compared to figure \ref{fig:scenarioIplot1} we have also translated the vertical axis from $\log(\lambda^2_{112})$ to $\log(g_{112}^2)$ using \eqref{eq:CtoG}, so the orange extrapolations are now approximately constant rather than sloping down. We fitted this rescaled data with a quadratic polynomial in $\Delta^{-1}$ to obtain the red points projected on the back surface. These constitute our final result, \emph{i.e.}, the red data points should be true upper bounds for $g_{112}^2$ in any flat-space S-matrix. Figure \ref{fig:scenarioIplot2} shows the same results more clearly.

Our data points in figure \ref{fig:scenarioIplot2} are in good agreement with the blue curve corresponding to the function
\be
\label{g112max}
(g_{112}^{\max})^2 = \frac{4 \left(\mu^2(4-\mu ^2)\right)^{3/2}}{|\mu ^2-2|}\,, \qquad \mu = m_2/m_1\,.
\ee
This curve is obtained from an analysis described in our companion paper \cite{paper2}, where we bootstrap two-dimensional scattering amplitudes directly. It can be obtained from the residue of the pole at $s = m_2^2$ of the two-dimensional S-matrix\footnote{More precisely, the residue is related to $(g_{112}^{\max})^2$ by a Jacobian factor ${\mathcal J}_2$ \cite{paper2}.}
\be
S(s) = \text{sgn}(m_2^2 - 2 m_1^2)\frac{\sqrt{s} \sqrt{4 m_1^2-s}+m_2\sqrt{4 m_1^2-m_2^2} }{\sqrt{s} \sqrt{4 m_1^2-s}-m_2 \sqrt{4 m_1^2-m_2^2}} \,. \la{SGmatrix}
\ee
As detailed in \cite{paper2}, we can \emph{prove} that this is the S-matrix that maximizes $g_{112}^2$ under the assumptions of scenario I. We find the agreement between our numerical data and \eqref{g112max} quite remarkable. In particular, neither the symmetry of \eqref{g112max} under $\mu^2 \to 4 - \mu^2$ nor the singularity at $\mu^2 = 2$ are in any way obvious from the setup of the CT problem and instead are an \emph{output} of our numerical analysis.\footnote{Preliminary numerical results indicate that the peak has finite height for all finite $\Delta$, so the divergence likely only occurs in the $\Delta \to \infty$ limit.}

Let us recap. Using the conformal bootstrap methods for the CT observables that correspond to a QFT in AdS we were able to obtain nonperturbative upper bounds on the residues $g_{112}^2$ for any flat-space QFT in two spacetime dimensions. Within numerical errors, these bounds are in agreement with the bounds obtained from a direct analysis of the flat-space S-matrix. We believe that this lends significant credibility to the relation between CT observables and the flat-space S-matrix.

\begin{figure}[t]
\begin{center}
\includegraphics[width=14cm]{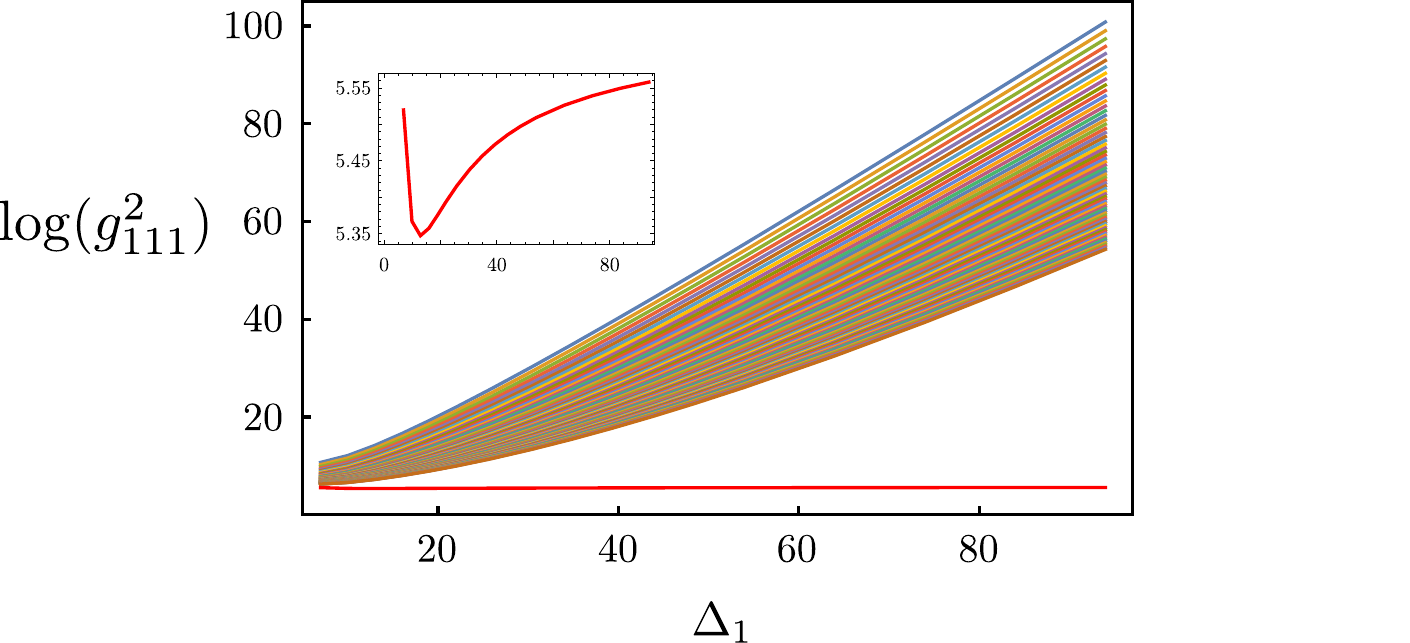}
\caption{Upper bound on $\log(g_{111}^2)$ as a function of $\Delta_1$, for a gap $\Delta_b = 1.85 \Delta_1$ in scenario II.
The different curves correspond to different numbers of derivatives $N$ ranging from 40 (top) to 300 (bottom). Each curve is an upper bound, which gets stronger as $N$ increases. In red we show the extrapolation to infinite $N$, enlarged in the inset. It varies relatively little and seems to asymptote to a constant at large $\Delta_1$.}
\label{fig:logope}
\end{center}
\end{figure}

\subsection{Scenario II}
In the previous subsection we presented the core ideas behind the numerical analysis and demonstrated the feasibility of the method for QFTs captured by scenario I. In this section we instead take a more in-depth look and aim for a precision analysis, this time in the context of scenario II. We recall that in this scenario we maximize $\lambda_{111}^2$ subject to the constraint that other operators have scaling dimensions greater than some value $\Delta_b$.

In figure \ref{fig:logope} we show the raw numerical bootstrap bounds as a function of $\Delta_1$, using a representative value $\Delta_b = 1.85 \Delta_1$. Other values of $\Delta_b/\Delta_1$ give similar results.  In contrast to figure \ref{fig:scenarioIplot1} we have chosen here to show the results directly for $g_{111}$ which we recall is related to the OPE coefficient $\lambda_{111}$ via equation \reef{eq:CtoG}. The different curves correspond to different values of $N$.

As before, we observe that our bounds vary substantially with $N$. We therefore performed an extrapolation to $N = \infty$, which should be free of artefacts due to finite computational resources and therefore a closer representative of actual physics. Aiming for the highest possible precision, we have in this case obtained data for values of $N$ up to 300 and subsequently fitted our best 30 results to a degree 29 polynomial in $N^{-1}$. The reason why we can get away with such an extreme fit is that our numerical bounds were obtained with a very small relative accuracy of $10^{-100}$. By doing various cross-checks we convinced ourselves of the reliability of this extrapolation procedure. Some more details on the extrapolation are provided below.

\begin{figure}[t]
\begin{center}
\includegraphics[width=14cm]{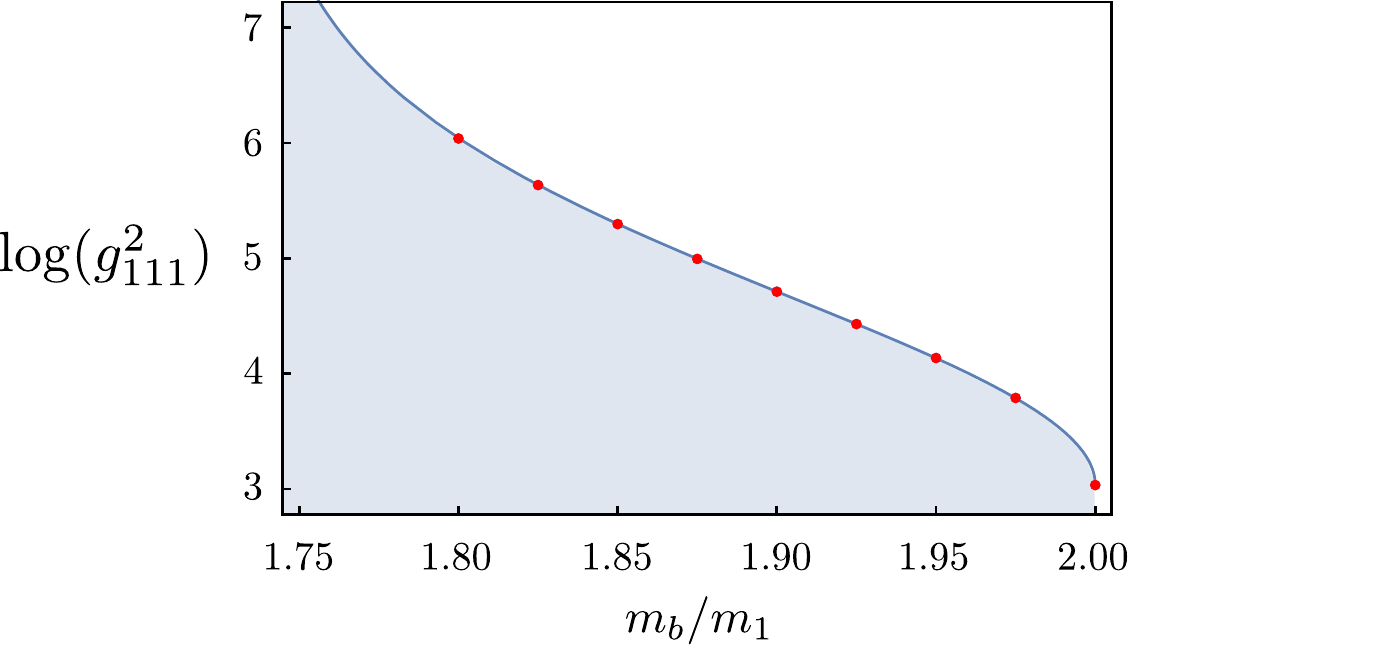}
\caption{Upper bound on the flat space self-coupling $g_{111}$ as a function of the mass ratio $m_b/m_1$. The dots are our numerical results obtained by extrapolation to infinite derivatives and infinite $\Delta_1$. The solid blue curve shows the same quantities for the exact S-matrix explained in the main text.}
\label{fig:flatcoupling}
\end{center}
\end{figure}

We once more discover an excellent match between the extrapolated curve and the exponential decrease predicted by equation \reef{eq:FSLcubiccoupling} with $\alpha = 1$, which is visible in figure \ref{fig:logope} as a nearly flat result. This allows us to perform a secondary extrapolation to infinite $\Delta_1$ by fitting a quadratic polynomial in $1/\Delta_1$. The results of this secondary extrapolation are shown as the red data points in figure \ref{fig:flatcoupling}. We claim that these are upper bounds on $g_{111}$ as a function of the mass ratio $m_b/m_1$, valid for any unitary two-dimensional QFT described by scenario II. 

Our numerical analysis again matches an \emph{exact} S-matrix bound which is shown as the solid curve in figure \ref{fig:flatcoupling}. As explained in our companion paper \cite{paper2}, this curve corresponds to the coupling $(g_{111}^{\max})^2$ determined from the amplitude
\bea
S(s)= \frac{\sinh(\theta)+i\,\sin(\alpha_1)}{\sinh(\theta)-i \sin(\alpha_1)} \cdot \frac{\sinh(\theta)+i\,\sin(\alpha_2)}{\sinh(\theta)-i \sin(\alpha_2)}  \label{eq:exactS}
\eea
where $\cosh(\theta/2)=s/2$, $\cos(\alpha_1/2)=1/2$ and  $\cos(\alpha_2/2)=m_b/(2m_1)$. Concretely we obtain that
\bea
(g_{111}^{\max})^2=\frac{36+24\sqrt{3} \sin(\alpha_2)}{\sqrt{3}-2\,\sin(\alpha_2)}
\eea
and our numerical results match this curve with a difference smaller than a part in a thousand! This strongly suggests that our extrapolations are reliable and a precision analysis is possible using the CT framework.

\subsubsection*{Spectrum and phase shift} 

Precisely when the numerical OPE coefficient bound is saturated we can extract an approximate \emph{solution} to the crossing symmetry equations as a side result from the numerical analysis \cite{Poland:2010wg,El-Showk2013}. This gives us an approximate spectrum that we can compare against our flat-space intuitions and use to test the phase-shift formula.

\begin{figure}[t]
\begin{center}
\includegraphics[width=14cm]{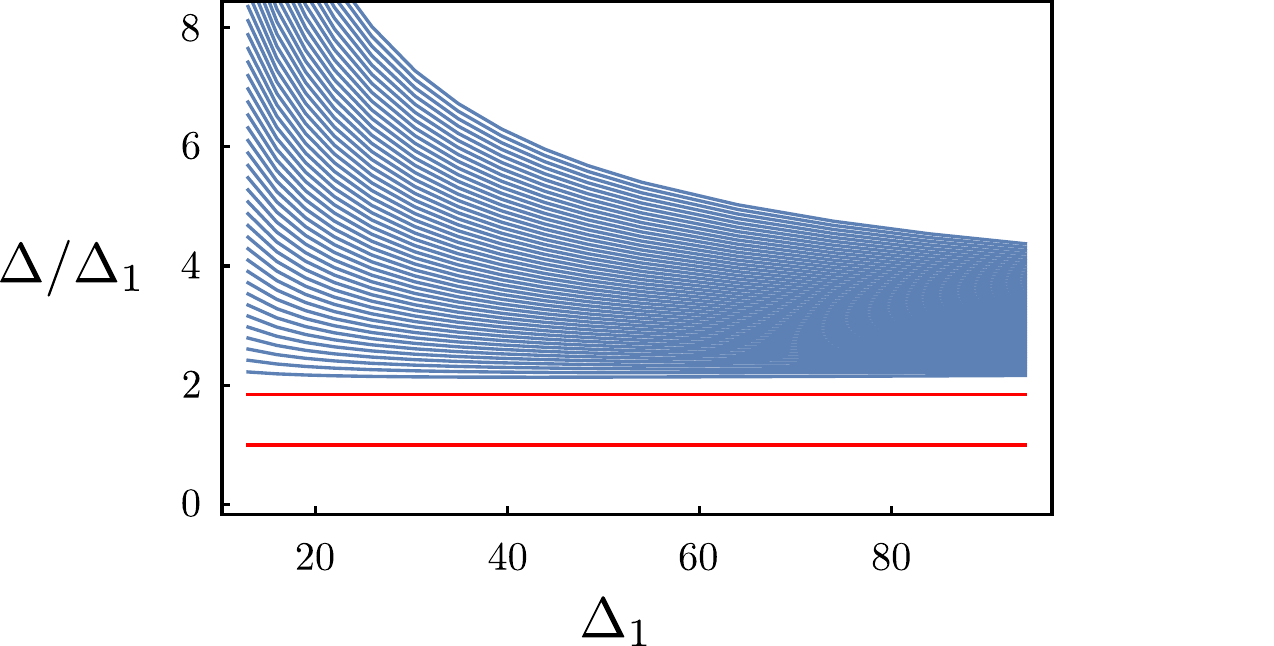}
\caption{Spectrum of operator dimensions of the solution to crossing symmetry that saturates the $N= 300$ bound for $\Delta_b/\Delta_1=1.85$, as a function of $\Delta_1$. As $\Delta_1$ increases the spectrum includes two bound states and a ``two-particle continuum'' starting at roughly $2\Delta_1$. 
}
\label{fig:S1spec}
\end{center}
\end{figure}

In figure \ref{fig:S1spec} we show the approximate \emph{spectrum} of the CT as obtained from our numerical analysis, for the specific case $\Delta_b/\Delta_1=1.85$ and for $N=300$. The figure shows a clear approach toward the spectrum of the flat-space amplitude \eqref{eq:exactS} for large $\Delta_1$, with no further operators in the gap between $1.85\Delta_1$ and $2\Delta_1$ and then a ``two-particle continuum'' above $2 \Delta_1$. Although we do not show it here, extrapolation to infinite $\Delta_1$ corroborates this picture.

\begin{figure}[t]
\begin{center}
\includegraphics[width=14cm]{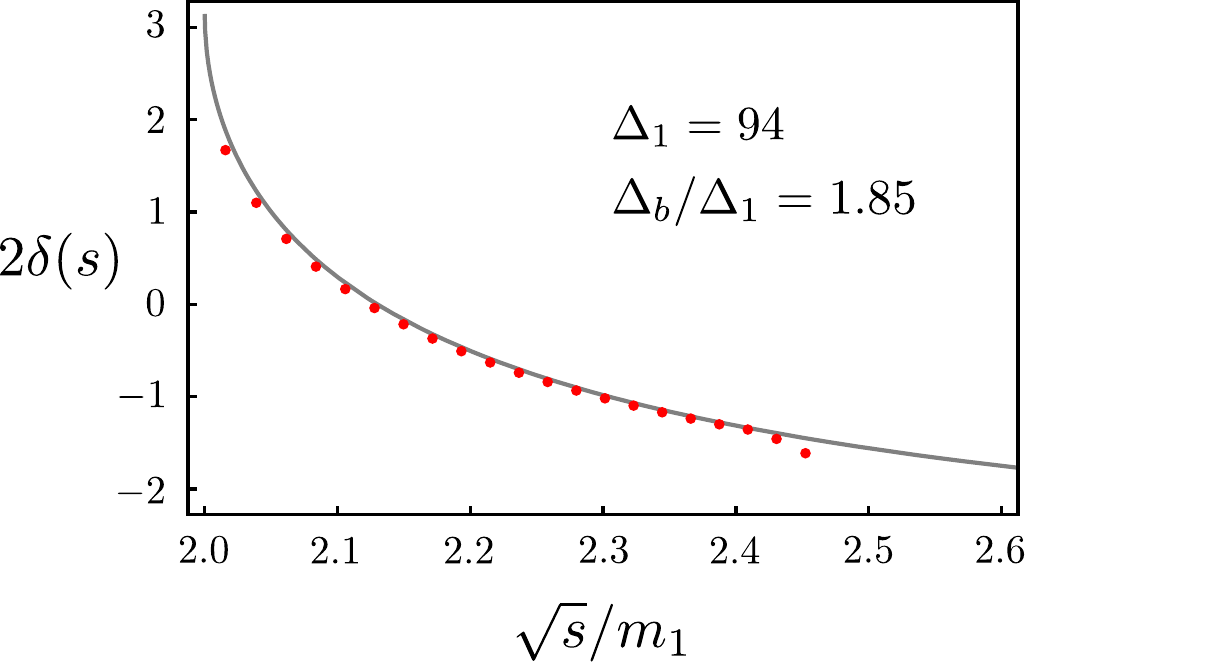}
\caption{Exact vs numerical phase shift. In black the exact phase shift corresponding to the S-matrix given in \reef{eq:exactS}. The red dots are determined numerically from the conformal bootstrap.
}
\label{fig:S1phase}
\end{center}
\end{figure}

We can go further and also compute the \emph{phase shift}. To do so we fix large $\Delta_1$ and extrapolate to infinite number of derivatives the spectrum of operator dimensions. In figure  \ref{fig:S1phase}, we plot the resulting phase shift, which is given by $e^{2i \delta(s)} = e^{-i \pi (\Delta-2\Delta_1)}$ for the discrete values of $\Delta/\Delta_1=\sqrt{s}/m_1$ that appear in the conformal block decomposition of the four-point function.
These results are compared with the phase shifts corresponding to the exact S-matrix $S(s)=e^{2i\delta(s)}$ given in \reef{eq:exactS}, and we see that the agreement is excellent.\footnote{For values of $\sqrt{s}/m_1$ greater than about 2.5 our numerical estimate of the spectrum is not reliable and we do not show this data here.}
Notice that our procedure corresponds to the application of formula (\ref{eq:phaseshiftformula}) with a small energy width $\delta E \sim 1$ so that there is only one primary operator per energy bin. This can seem surprising because formula (\ref{eq:phaseshiftformula}) was derived assuming $\delta E \gg 1$. Therefore, we should be able to obtain the same phase shift using $\delta E \gg 1$ which means many operators per energy bin. Fortunately this follows from the natural assumption that the limit $\Delta_1 \to \infty$ leads to a figure similar to \ref{fig:S1phase} with the primary operators (red dots) densely packed along the black curve. In that case, averaging $e^{i\pi(2\Delta_1-\Delta)}$ over all operators $\mathcal{O}_\Delta$  with $|\Delta-E|<\delta E\ll E$ gives a result independent of $\delta E$ in the flat space limit $E\sim \Delta_1 \to \infty$. 

\subsubsection*{Extrapolations}
\begin{figure}[t]
\begin{center}
\includegraphics[width=16cm]{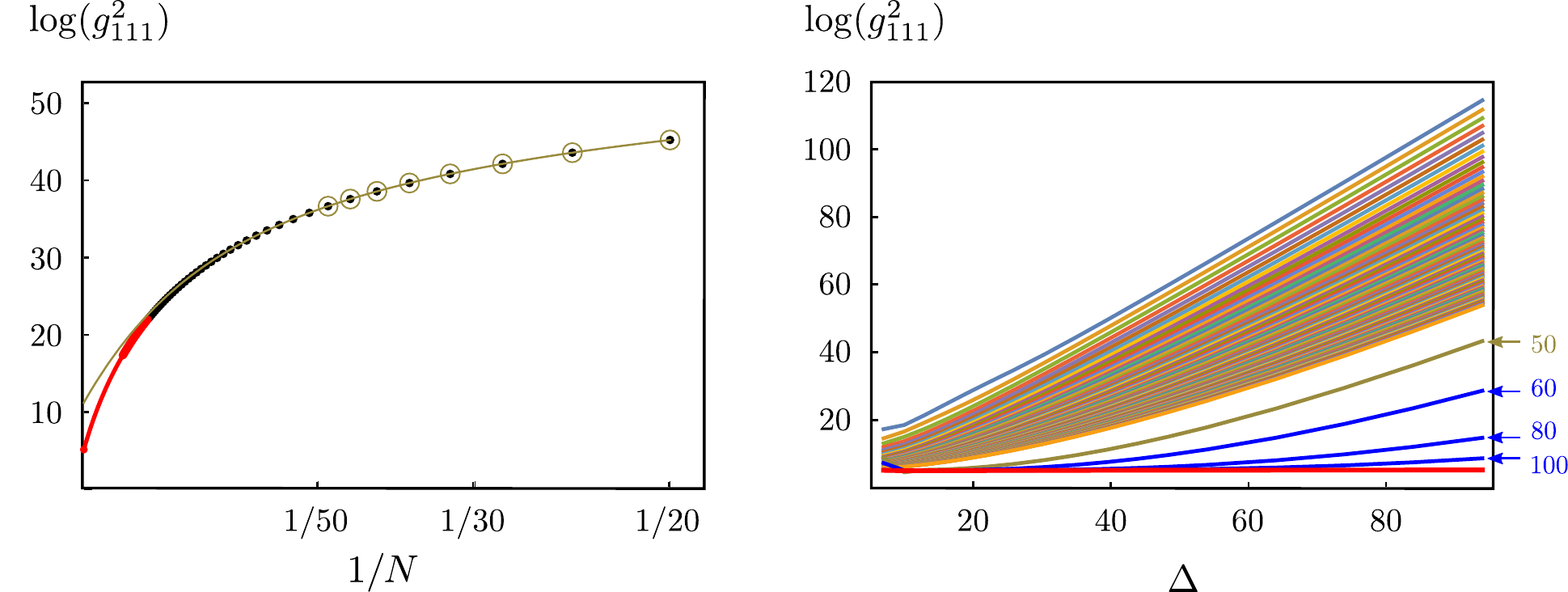}
\caption{
Testing the extrapolations to infinite $N$. See the main text for explanations.
}
\label{fig:extrapolationsdetails}
\end{center}
\end{figure}

Using figure \ref{fig:extrapolationsdetails} we will now provide a few more details concerning our $N \to \infty$ extrapolation procedure. We emphasize that there is currently no analytic understanding of the large $N$ behavior of numerical bootstrap bounds, so we will restrict ourselves to a qualitative discussion.\footnote{In other bootstrap analyses the extrapolations have nevertheless been very useful, see \emph{e.g.} \cite{Beem:2015aoa}, and yielded results that are consistent with expectations.}

On the left of figure \ref{fig:extrapolationsdetails} we show the result of a single extrapolation. The dots correspond to all our raw data for a (representative) data point with $\Delta_1 =  39.4$ and $\Delta_b/\Delta_1 = 1.85$. The red curve shows the extrapolation using a degree 29 polynomial in $N^{-1}$ that uses only the last 30 data points (also in red). This is the extrapolation that we used for all the scenario II results. Obviously, our $N \to \infty$ estimate is given by the intersection point of this curve with the vertical axis. We have also drawn the yellow curve, which is an indication of the kind of result that we would have obtained with fewer data points. For this particular example we used a degree seven fit through the eight circled data points with $20 \leq N \leq 50$. This extrapolation is quite a bit off, and we conclude that such values of $N$ are insufficient to obtain a reliable result. Qualitatively we can explain the unreliability of our low-$N$ extrapolations by the sharp downward slope of the raw data points as $N$ increases, which is not captured by the low $N$ values. This poses an obvious challange when the numerical results become more difficult to obtain, for example when we consider CTs in higher dimensions.

On the right we show, besides the raw data, the extrapolations involving all data points with $N$ ranging from $20$ to $50$, $60$, $80$ and $100$. For larger values of $\Delta$ we observe not merely a worse convergence of the numerical bounds but also far less reliable extrapolations. Clearly, only the last curve really comes close to our best extrapolations.

\section{Conclusion}
\label{sec:conclusions}
By putting a QFT in an AdS background we can define a set of boundary `conformal theory' observables which are near-identical to the correlation functions of a CFT; they only lack a stress tensor operator. Upon taking the AdS radius to infinity these observables should transform smoothly into the flat-space S-matrix of the QFT. This paper offers three concrete results that solidify this idea:
\begin{itemize}
	\item We proposed a precise formula for the map between the CT correlation functions and scattering amplitudes. Our formula fundamentally relies on the Mellin space description of the CT observables \cite{Mack2009,Mack2009a}, and we have shown that our formula works in specific perturbative examples.
	\item For physical energies we have also shown that the \emph{phase shift} can be obtained directly as a limit of CFT data. At weak coupling this matches a known result \cite{Fitzpatrick:2010zm} but we claim that it holds nonperturbatively.
	\item We have applied numerical conformal bootstrap methods to the CT observables for two-dimensional unitary QFTs, and by means of various extrapolations obtained nonperturbative bounds for their flat-space scattering amplitudes. These results match precisely to the analytic S-matrix bootstrap discussed in \cite{paper2}.
\end{itemize}
Our results highlight once more the remarkable richness of the conformal crossing symmetry equations, which apparently ``know'' not only about CFTs but also about massive QFTs in AdS. Furthermore, we have shown above that the modern bootstrap methods of \cite{Rattazzi:2008pe} allow us to successfully extract this information and translate it into precise upper bounds. Our results clearly raise the urgent question whether similar nonperturbative results can be obtained for higher-dimensional theories. In that case the numerical analysis is more involved: there are spinning operators in the CT and there are two cross-ratios rather than a single one. We nevertheless expect to report on this question in the near future. Another avenue for progress would be the numerical analysis of multiple correlators as in \cite{Kos:2014bka}. This has the potential to drastically improve our numerical bounds and, as we explain in \cite{paper2}, is likely to be essential for the generalization of the S-matrix bounds to non-integrable theories.

To gain a better intuition into the structure of the CT observables it would be interesting to work out further explicit examples. We can for example consider weakly coupled theories. Even in this case there are numerous subtleties that arise when placing quantum field theories in hyperbolic space, mostly related to the matter of boundary conditions. Already in the simplest example, namely a free scalar field, the work of Breitenlohner and Freedman shows that not all boundary conditions are consistent with general positivity conditions. Infrared divergences introduce further complications: for classical field theories this has been demonstrated convincingly already in older work on holographic renormalization \cite{Skenderis:2002wp} in the context of AdS/CFT, whereas for loop diagrams some work remains to be done. For Yang-Mills theories it would further be interesting to understand the space of boundary conditions better \cite{Witten:2003ya,Aharony:2010ay}. Finally there are some questions about confining theories in hyperbolic space, discussed already in \cite{Callan1990} and more recently in \cite{Aharony:2012jf}. Although we have sidestepped these and other subtleties in this work, they certainly deserve further attention.

It is not hard to check that the S-matrices that we recover at our numerical bounds saturate unitarity without particle production \cite{paper2}. Therefore, in the cases that they correspond to a physical theory this is bound to be an integrable QFT. This naturally raises the question whether integrability survives in some form when the QFTs are put in hyperbolic space, which to the best of our knowledge is currently unanswered. It would be great to have integrable strongly coupled boundary CTs as analytic examples where we can explicitly recover the flat-space S-matrices from the formulae we presented above. One example is to consider the integrable massive deformation
of the Ising model that just corresponds to giving mass to the free
fermions. In this case, we should be able to start from any of the
3 possible BCFT of the 2D Ising in the UV, and see where we end up
in the IR. This has been studied in \cite{Doyon:2003nb,Doyon:2004fv}.

Our viewpoint could also be of use for the analytic properties of the S-matrix in QFT. These are usually \emph{assumed} to be relatively straightforward, with simple poles and cuts as dictated by a perturbative analysis, but we are not aware of any nonperturbative proof. In contrast, the analyticity of the boundary CT observables follows essentially from the operator-state correspondence and is therefore on a much firmer footing. The fact that our CT analysis agrees with the S-matrix bootstrap which fundamentally assumes analyticity is remarkable. Perhaps our viewpoint could be used to \emph{define} the famous ``analytic'' S-matrix as the flat-space limit of the boundary CT observables. In this way, the analyticity properties of the S-matrix would follow from the well established meromorphicity of the Mellin amplitude.

In this paper we focused on the flat-space limit which practically implied sending $\Delta \to \infty$. However we need not have done so: the CT construction shows that the undoubtedly rich physics of QFTs in AdS is described by finite values of $\Delta$.
In such cases there exists a one-parameter family of CTs corresponding to each relevant bulk coupling $\mu_i R$.
These lines of CTs begin (and possibly end) at BCFTs corresponding to the UV (and possibly IR) bulk CFT, but in between they describe the physics of massive theories. It would of course be very interesting to get a handle on such flows, and we may even speculate that they are sometimes described by ``extremal flows'' as in \cite{El-Showk:2016mxr}. In any case, we find it striking that the $d$ dimensional crossing symmetry equations know about about $d+1$ dimensional massive QFT physics in such a crisp way. What else do they know?

\section*{Acknowledgments}
We would like to thank the GGI for hospitality during the final stages of this work, and the participants of the program ``Conformal Field Theories and Renormalization Group Flows in Dimensions $d>2$'' for numerous interesting discussions. 
Research at the Perimeter Institute is supported in part by the Government of Canada through NSERC and by the Province of Ontario through MRI. 
MFP is supported  by  a  Marie  Curie Intra-European  Fellowship  of  the  European  Community's  7th Framework Programme under contract number PIEF-GA-2013-623606. 
\emph{Centro de F\'{i}sica do Porto} is partially funded by FCT. 
The research leading to these results has received funding from the People Programme (Marie Curie Actions) of the European Union's Seventh Framework Programme FP7/2007-2013/ under REA Grant Agreement No 
 317089 (GATIS) and the grant CERN/FIS-NUC/0045/2015.
 JP is supported by the National Centre of Competence in Research SwissMAP funded by the
Swiss National Science Foundation.

\appendix

\section{RG flows in hyperbolic space}
\label{appendix:rgflow}
We will consider an RG flow connecting a CFT$_{UV}$ to a CFT$_{IR}$ or to a gapped phase. 
Let us start by placing the CFT$_{UV}$ in hyperbolic space. This requires the choice of conformal invariant boundary conditions. In fact, since hyperbolic space is conformal to half of Euclidean space, 
this is equivalent to Boundary CFT (BCFT). 
For example, correlators of local primary operators $\phi_i$ of the CFT$_{UV}$ in AdS are simply related to the same correlators in flat space BCFT, 
\[
\left\langle \phi_{1}(z_1,x_{1})\dots \phi_{n}(z_n,x_{n})\right\rangle_{AdS_{d+1}} =
z_1^{\tilde{\Delta}_1}\dots z_n^{\tilde{\Delta}_n}
\left\langle \phi_{1}(z_1,x_{1})\dots \phi_{n}(z_n,x_{n})\right\rangle_{\mathbb{R}^d\times\mathbb{R}^+}
\]
where $\tilde{\Delta}_i$ is the UV scaling dimension of  $\phi_i$.
Furthermore, the boundary operators $\mathcal{O}_k$ defined by the operator boundary expansion (\ref{AdSOBE}) are just the standard boundary operators of BCFT.

We then turn on a relevant deformation of the bulk CFT$_{UV}$.
Formally, we can write the boundary correlators as follows
\[
G_{1\dots n}(x_1,\dots,x_n;\mu R) =
\frac{
\left\langle       \mathcal{O}_{1}(x_{1})\dots \mathcal{O}_{n}(x_{n})   e^{ \mu^{D-\tilde{\Delta}_{r}}\int_{AdS}d^{D}x\phi_{r}(x)}\right\rangle _{AdS}}{\left\langle e^{ \mu^{D-\tilde{\Delta}_{r}}\int_{AdS}d^{D}x\phi_{r}(x)}\right\rangle _{AdS}}\,,
\]
where $\phi_r$ is a relevant scalar operator of the bulk CFT$_{UV}$ with dimension $\tilde{\Delta}_{r}<D$. The mass scale $\mu$ in AdS gives rise to a dimensionless parameter $\mu R$ that characterizes the boundary correlators along the flow. In particular, the spectrum of boundary scaling dimensions $\Delta_k$ will vary continuously with the parameter $\mu R$. If the RG flow ends in a CFT$_{IR}$, then $ \Delta_k(\mu R)$ interpolates between the spectrum of boundary operators of the BCFTs describing the UV and IR fixed points in AdS, when $\mu R$ varies from $0$ to $\infty$.   
If the RG flow ends in a gapped phase, then all boundary dimensions $\Delta_k(\mu R)$ become parametrically large when $\mu R \to \infty$ and one can read off the mass spectrum of the bulk QFT from the limit 
\beq
\frac{m_k}{m_1}=\lim_{\mu R \to \infty} \frac{\Delta_k(\mu R)}{\Delta_1(\mu R)}\,.
\eeq

One simple example is the flow of a free scalar field when we turn
on the relevant deformation corresponding to the mass. 
The UV starting point
depends on the boundary condition we choose for the free scalar. There
are two possible BCFT: Dirichlet, which has $\phi=0$ at the boundary
and Neumann, which has $\partial_z\phi=0$ at the boundary. If
we choose the Dirichlet BCFT, then the lowest boundary operator is $ \mathcal{O}(x)= \partial_z\phi(0,x)$
with dimension $\Delta=\frac{d+1}{2}$, where $d=D-1$ is the boundary
dimension. If we choose the Neumann BCFT, then the lowest boundary operator is $\mathcal{O}(x)=\phi(0,x)$
with dimension $\Delta=\frac{d-1}{2}$. These are the two possible
values of the dimension of the CFT operator dual to a scalar in AdS$_{d+1}$
with mass squared given by $-\frac{d^{2}-1}{4R^2}$, which is the mass
of a conformally coupled scalar. When we turn on the relevant deformation $\frac{1}{2}\mu^2\phi^2$,
we find
\begin{equation}
\Delta=\frac{d\pm\sqrt{1+(2\mu R)^{2}}}{2}\ .\label{eq:DeltaFreeScalar}
\end{equation}
This means that if we start with the Neumann BCFT we can not increase
$\mu R$ arbitrarily without violating the unitary bounds of the BCT.
On the other hand, starting from the Dirichlet BCFT we can go all
the way into the deep infrared to find $ \Delta \approx\mu R$,
as expected for a particle of mass $\mu\ll \frac{1}{R}$ in AdS.

This example shows that not all possible boundary conditions of the CFT$_{UV}$ are convenient to study RG flows. It would be interesting to understand this point in more detail. However, for our conformal bootstrap approach we only have to assume that there is at least one boundary condition that is consistent along the entire flow.

\subsection{Stress-energy tensor}
\label{subapp:stresstensor}
The bulk QFT has a local stress-energy tensor. Here, we would like to discuss what boundary operators can be obtained by pushing the bulk stress tensor to the boundary of AdS. 
The asymptotic expansion of the stress tensor 
must be compatible with the conservation equations
\[
z^{d}\partial_{z}\left(z^{-d}T_{\ z}^{z}\right)+\partial_{\mu}T_{\ z}^{\mu}+\frac{1}{z}T_{\ \mu}^{\mu}=0\ ,\qquad z^{d+1}\partial_{z}\left(z^{-d-1}T_{\ \nu}^{z}\right)+\partial_{\mu}T_{\ \nu}^{\mu}=0\ .
\]
This suggests the following behaviour
\begin{align}
&T_{\ z}^{z}\approx z^{\Delta_{\mathcal{D}}}\mathcal{D}\ ,\qquad T_{\ \nu}^{\mu}\approx z^{\Delta_{t}}t_{\ \nu}^{\mu}-\frac{\Delta_{\mathcal{D}}-d}{d}z^{\Delta_{\mathcal{D}}}\delta_{\nu}^{\mu}\mathcal{D}\ ,
\\&T_{\ \nu}^{z}\approx\frac{1}{d}z^{\Delta_{\mathcal{D}}+1}\partial_{\nu}\mathcal{D}-\frac{1}{\Delta_{t}-d}z^{\Delta_{t}+1}\partial_{\mu}t_{\ \nu}^{\mu}\ ,
\end{align}
where $\mathcal{D}$ is a scalar boundary operator, $t_{\mu\nu}$ is a spin 2 (traceless) boundary operator and we neglected the contribution from higher dimension operators.
The boundary scaling dimensions $\Delta_{\mathcal{D}}$ and
$\Delta_{t}$ can vary independently along the flow as we increase $\mu R$.
In the CFT$_{UV}$ the bulk stress tensor is traceless. This gives $T_{\ z}^{z}+T_{\ \mu}^{\mu} =
(d+1-\Delta_{\mathcal{D}}) \mathcal{D}=0$, which implies that  $\Delta_{\mathcal{D}}=d+1$ and $\mathcal{D}$ is called the displacement operator in BCFT.

Let us check these equations explicitly in the case of a massive free scalar
field in AdS. The stress tensor is given by (see for example \cite{Osborn:1999az})
\[
T_{\ b}^{a}=\nabla^{a}\phi\nabla_{b}\phi-\frac{1}{4d}\left[(d-1)\nabla^{a}\nabla_{b}+\delta_{b}^{a}\left(\nabla^{2}+d(d-1)\right)\right]\phi^{2}\,,
\]
and the equation of motion is
\[
\nabla^{2}\phi=-\frac{d^{2}-1}{4}\phi+\mu^{2}\phi\,.
\]
One can check conservation $\nabla_{a}T_{\ b}^{a}=0$ and $T_{\ a}^{a}=-\mu^{2}\phi^{2}$.
The asymptotic behavior of the bulk scalar field  is given by
\[
\phi(z,x)\approx z^{ \Delta } \mathcal{O} (x)\,,
\]
where the scaling dimension is given in (\ref{eq:DeltaFreeScalar}).
This gives
\[
T_{\ z}^{z}\approx\frac{2 \Delta -d+1}{4}z^{2 {\Delta}} {\mathcal{O}}^{2}\ ,\qquad\Delta_{t}=2 {\Delta}+2\ .
\]
For Dirichlet boundary conditions, $2 {\Delta}$ starts from
$d+1$ in the UV and grows after the massive deformation. In this
case, $\Delta_{\mathcal{D}}=2 {\Delta}$ and the lowest boundary spin
2 operator has   $\Delta_{t}=\Delta_{\mathcal{D}}+2$.
With Neumann boundary conditions, the situation is more subtle. In this case the boundary operator $\mathcal O$ has dimension $\Delta=\frac{d-1}2$ in the CFT$_{UV}$. 
In agreement with the discussion above, one can check that $\mathcal D\sim (\partial \mathcal O)^2$ and  $\Delta_{\mathcal{D}}=d+1$ as required for a displacement operator.
However,  as soon as we move away from the UV fixed point, the stress-tensor is no longer required to be traceless and this allows for a coupling to the scalar operator $\mathcal O^2$ which has (smaller) dimension $2\Delta=d-1+O(\mu^2R^2)$.

\section{Scattering states \label{sec:scatteringstates}}

Let us start with the case of AdS$_2$ and consider the following state in radial quantization
\be
|\psi \rangle = \int_0^1 dy\, 4y\left(1-y^2\right)^{\Delta-2} \mathcal{O}_\Delta (y)|0\rangle
=\sum_{n=0}^\infty \frac{2 \Gamma (\Delta -1) \Gamma
   \left(\frac{n}{2}+1\right)}{n!
   \Gamma
   \left(\frac{n}{2}+\Delta
   \right)} \partial^n\mathcal{O}_\Delta (0)|0\rangle\,.
\ee
We are interested in the case where the primary state $\mathcal{O}_\Delta (0)|0\rangle$ is the lowest energy state of a stable particle in AdS$_2$. Then $|\psi \rangle $ is just a specific linear combination of boosted versions of this one-particle state.
The reason for this particular choice becomes clear once we consider the associated bulk wave-function,
\be
\psi(\tau,\rho)\propto \int_0^1 dy\, 4y\left(1-y^2\right)^{\Delta-2}
\left(
\frac{e^\tau \cos \rho }{e^{2\tau} \cos^2 \rho + \left(e^\tau \sin \rho-y\right)^2}
\right)^\Delta
\label{bulkpsieuclidean}
\ee 
where the last factor is the scalar bulk to boundary propagator written in Euclidean bulk global coordinates. We want to study the Lorentzian time evolution of this state. In particular, we want to focus on a small flat space scattering region   $t\sim \rho \sim \frac{1}{\Delta} \ll 1$ where the Lorentzian time $t$ is given by
$\tau \to i( t+\frac{\pi}{2})$. In other words, the scattering event will happen after a time interval of $\frac{\pi}{2}$ as depicted in figure \ref{fig:Phaseshift}.
In this small flat space region, we have
\be
\psi\sim \int_0^1 dy\, \frac{4y}{\left(1-y^2\right)^{2}}
e^{-i\Delta t \frac{1+y^2}{1-y^2} -i\Delta \rho \frac{2y}{1-y^2}  }\ .
\ee
Changing to the integration variable $\omega=\Delta  \frac{1+y^2}{1-y^2}$, we obtain
\be
\psi\sim \int_\Delta^\infty d\omega
e^{-it \omega  -i \rho \sqrt{\omega^2-\Delta^2}    }\ ,
\ee
which is a linear combination of all on-shell states with energy varying from $\Delta$ to $\infty$ and with negative spatial momentum. The important feature of this state is that the spectral weight is constant. This makes it easy to construct localized wave packets by considering  projections to an  energy band of width $\delta \omega $ satisfying  $\Delta \gg \delta \omega \gg 1$. 
More precisely, the wave packet 
 \be
\psi_q(t,\rho)\sim \int_\Delta^\infty d\omega \,q(\omega)
e^{-it \omega  -i \rho \sqrt{\omega^2-\Delta^2}    }\ ,
\ee
where $q(\omega)$ is a smooth envelope of width $\delta \omega $ around a central frequency $E\sim\Delta$,  corresponds to the state
\be
|\psi_q\rangle = \int_0^1 dy\,q\left( \Delta  \frac{1+y^2}{1-y^2}\right) \, 4y\left(1-y^2\right)^{\Delta-2} \mathcal{O}_\Delta (y)|0\rangle\ .
\ee
It is instructive to consider the wave function of this state at $\tau=0$.
Using the same logic as in \eqref{bulkpsieuclidean}, we find
\be
\psi_q(0,\rho)\sim \int_0^1 dy\,q\left( \Delta  \frac{1+y^2}{1-y^2}\right)\, 4y\left(1-y^2\right)^{\Delta-2}
\left(
\frac{  \cos \rho }{  \cos^2 \rho + \left(  \sin \rho-y\right)^2}
\right)^\Delta\ .
\ee
For positive $\rho$, the integral is dominated by a saddle point at $y= \tan \rho$ and we obtain
\be
\psi_q(0,\rho)\sim \frac{ \cos \rho}{1+\cos \rho}
q\left( \Delta \tan \rho \right)\ .
\ee
The means that the initial wave function is peaked at $\rho=\arctan \frac{E}{\Delta}$ with a width   $\delta\rho \sim \delta \omega /\Delta \ll 1$.
For negative $\rho$ the initial wave-function is exponentially small.

The next step is to construct scattering states. 
The natural starting point is
\be
|\Psi \rangle =\left[ \int_0^1 dy_1\int_{-1}^0 dy_2\, 16y_1y_2\left(1-y_1^2\right)^{\Delta_1-2}
\left(1-y_2^2\right)^{\Delta_2-2} \mathcal{O}_1 (y_1)  \mathcal{O}_2 (y_2)|0\rangle
\right]_{primaries}
\label{AdS2stateallenergies}
\ee
projected to primary states so that there is no center of mass motion.
Using the OPE 
\be
\mathcal{O}_1 (y_1)  \mathcal{O}_2 (y_2) =  \sum_{\Delta} \lambda_{\Delta}
(y_1-y_2)^{\Delta-\Delta_1-\Delta_2} \left[ \mathcal{O}_\Delta(0)+descendants \right]\,,
\ee
we find
\be
|\Psi \rangle = \sum_{\Delta} \bar{w}(\Delta) \lambda_{\Delta} \mathcal{O}_\Delta(0)|0 \rangle\,,
\ee
where
\be
\bar{w}(\Delta) = \int_0^1 dy_1\int_{-1}^0 dy_2\, 16y_1y_2\left(1-y_1^2\right)^{\Delta_1-2}
\left(1-y_2^2\right)^{\Delta_2-2}(y_1-y_2)^{\Delta-\Delta_1-\Delta_2}\ .
\ee
For large $\Delta \sim \Delta_1 \sim \Delta_2 \gg1$ this integral is dominated by a saddle point at
\be
y_1^\star = \sqrt{\frac{(\Delta-\Delta_1-\Delta_2)(\Delta-\Delta_1+\Delta_2)}{(\Delta+\Delta_1+\Delta_2)(\Delta+\Delta_1-\Delta_2)}}\,,\qquad
y_2^\star = - \sqrt{\frac{(\Delta-\Delta_1-\Delta_2)(\Delta+\Delta_1-\Delta_2)}{(\Delta+\Delta_1+\Delta_2)(\Delta-\Delta_1+\Delta_2)}}\ ,\nonumber
\ee
where we assumed that the total energy $\Delta> \Delta_1+\Delta_2$. 
It is nice to check that this saddle corresponds to the total energy
\be
\omega_1+\omega_2 = \Delta_1  \frac{1+(y_1^\star)^2}{1-(y_1^\star)^2}+
\Delta_2 \frac{1+(y_2^\star)^2}{1-(y_2^\star)^2} = \Delta
\ee
and to zero total spatial momentum
\be
\sqrt{\omega_1^2-\Delta_1^2}-\sqrt{\omega_2^2-\Delta_2^2} = \Delta_1  \frac{2y_1^\star }{1-(y_1^\star)^2}
+
\Delta_2 \frac{2y_2^\star}{1-(y_2^\star)^2} = 0\ .
\ee
This means that the normalized state
\be
|\Psi (E) \rangle = \frac{1}{\sqrt{N(E)}}\sum_{|\Delta-E|<\delta E} \bar{w}(\Delta) \lambda_{\Delta} |\Delta\rangle\,,\qquad
N(E)=\sum_{|\Delta-E|<\delta E} \left[\bar{w}(\Delta) \lambda_{\Delta}\right]^2\,,
\ee
with $E>\Delta_1+\Delta_2 \gg \delta E \gg 1$, is an appropriate scattering state.
Notice that the slow dependence (power law) of $\bar{w}$ on $\Delta$ cancels out in this state because the energy band $\delta E$ is much smaller than the average energy $E$. Therefore, it is sufficient to use the exponential dependence
\be
\bar{w}(\Delta) \to w(\Delta)\equiv \left[
\frac{4\Delta^2 (\Delta-\Delta_1-\Delta_2) }{
(\Delta^2-\Delta_{12}^2)(\Delta+\Delta_1+\Delta_2) }
\right]^{\frac{\Delta}{2}}
\frac{
\left(\frac{\Delta-\Delta_{12}}{\Delta+\Delta_{12}}\right)^{\frac{\Delta_{12}}{2}} }{
\left[ \Delta^2 -(\Delta_1+\Delta_2)^2 \right]^{\frac{\Delta_1+\Delta_2}{2}} }
\ ,
\ee
where $\Delta_{12}=\Delta_1-\Delta_2$.

The generalization to higher spacetime dimensions is straightforward. 
We can start from a state analogous to \eqref{AdS2stateallenergies} by placing the operators at points $y_1 \hat{n}$ and $y_2 \hat{n}$ for some unit vector $\hat{n}$. Then, we project to primaries of a given spin $l$ and with scaling dimension $\Delta$ in an energy band $|\Delta-E|<\delta E$. This gives
\be
|\Psi_l (E) \rangle = \frac{1}{\sqrt{N_l(E)}}\sum_{|\Delta-E|<\delta E} w(\Delta) \lambda_{\Delta,l} 
|\Delta,l\rangle\,,\qquad
N_l(E)=\sum_{|\Delta-E|<\delta E} \left[ w(\Delta) \lambda_{\Delta,l}\right]^2\,,
\ee
where $|\Delta,l\rangle = \hat{n}_{\mu_1} \dots \hat{n}_{\mu_l} \mathcal{O}_{\Delta,l}^{\mu_1\dots \mu_l}(0) |0\rangle$.

\section{Scattering amplitudes from Mellin amplitudes\label{sec:apendix:TfromM}}

In this appendix, it will be convenient to use the embedding formalism \cite{Dirac1936} where a point in AdS is represented by a vector $X\in \mathbb{R}^{d+1,1}$ such that $X\cdot X=-R^2$. A boundary point is represented by a null ray $P\sim \lambda P$ ($\lambda \in \mathbb{R}$) for $P\in  \mathbb{R}^{d+1,1}$ and $P\cdot P=0$.

Mellin amplitudes $M(\gamma_{ij})$ are defined by \cite{Mack2009,Mack2009a}
\[
\left\langle \mathcal{O}_{1}(P_{1})\dots\mathcal{O}_{n}(P_{n})\right\rangle =\int[d\gamma]M(\gamma_{ij})\prod_{1\le i<j\le n}\frac{\Gamma(\gamma_{ij})}{\left(-2P_{i}\cdot P_j\right)^{\gamma_{ij}}}\,,
\]
where the Mellin variables obey the constraints
\begin{equation}
\gamma_{ij}=\gamma_{ji}\ ,\qquad\gamma_{ii}=-\Delta_{i}\ ,\qquad\sum_{i=1}^{n}\gamma_{ij}=0\ .\label{eq:Mellinconstraints}
\end{equation}
From the Mellin amplitude, the associated flat space scattering amplitude
can be obtained from formula (\ref{eq:FSLgeneralformula}) which we rewrite here in the slightly different way 
\begin{equation}
(m_{1})^{a}\,T(k_{i})=\lim_{\Delta_{1}\to\infty}\frac{(\Delta_{1})^{a}}{\mathcal{N}}M\left(\gamma_{ij}= \frac{\Delta_{i}\Delta_{j}+R^2 k_{i}\cdot k_{j}}{\Delta_{1}+\dots+\Delta_{n}} +O(\Delta_1)^0 \right)\label{eq:FSLformula}
\end{equation}
where the AdS radius $R$ appears in the relation $m_i^2 R^2 =\Delta_i (\Delta_i-d)$. 
For completeness, we also reproduce here the normalization factor
\[
\mathcal{N}=\frac{1}{2}\pi^{\frac{d}{2}}\Gamma\left(\frac{\sum\Delta_{i}-d}{2}\right)\prod_{i=1}^{n}\frac{\sqrt{\mathcal{C}_{\Delta_{i}}}}{\Gamma(\Delta_{i})}\,.
\]
In this appendix we shall test this formula in some simple examples, present a perturbative derivation and analyse its consequences for the analytic structure of the flat space S-matrix.
Before that, notice that in   formula (\ref{eq:FSLformula}) the Mellin variables take values consistent
with the constraints (\ref{eq:Mellinconstraints}). More precisely, the parameterization holds for $\gamma_{ij}$ with $i\neq j$ ($\gamma_{ii}$ should still be set to $-\Delta_i$ explicitly), and by adding the finite piece
\beq
\frac{d}{n-2}\left[\frac{\Delta_{i}+\Delta_{j}}{\Delta_{1}+\dots+\Delta_{n}}-\frac{1}{n-1}\right]
\eeq
we guarantee consistency with the last constraint in \reef{eq:Mellinconstraints}.

\subsection{Examples}

We start by testing formula  (\ref{eq:FSLformula})   with simple Witten diagrams. 

\subsubsection{Contact interaction}

Consider the simplest contact interaction $g\phi_{1}\dots\phi_{n}$
in AdS. This leads to a boundary $n$-point function
\[
\left\langle \mathcal{O}_{1}(P_{1})\dots\mathcal{O}_{n}(P_{n})\right\rangle =g\int_{AdS}dX\prod_{i=1}^{n}\frac{R^{\frac{1-d}{2} }\sqrt{\mathcal{C}_{\Delta_{i}}}}{(-2P_{i}\cdot X/R)^{\Delta_{i}}}
\]
where we are using the embedding formalism and normalized the boundary
operators to have unit two point function. This gives the Mellin amplitude
\[
M=\frac{1}{2}gR^{n\frac{1-d}{2}+d+1}\pi^{\frac{d}{2}}\Gamma\left(\frac{\sum\Delta_{i}-d}{2}\right)\prod_{i=1}^{n}\frac{\sqrt{\mathcal{C}_{\Delta_{i}}}}{\Gamma(\Delta_{i})}
\]
and through formula (\ref{eq:FSLformula}) we obtain the scattering
amplitude $T=g$. Notice that the coupling $g$ is dimensionful and
the powers of the AdS radius $R$ appearing in the Mellin amplitude
make the combination dimensionless.

\subsubsection{Scalar exchange\label{sub:Scalar-exchange}}

\begin{figure}
\begin{center}
\includegraphics[width=16cm]{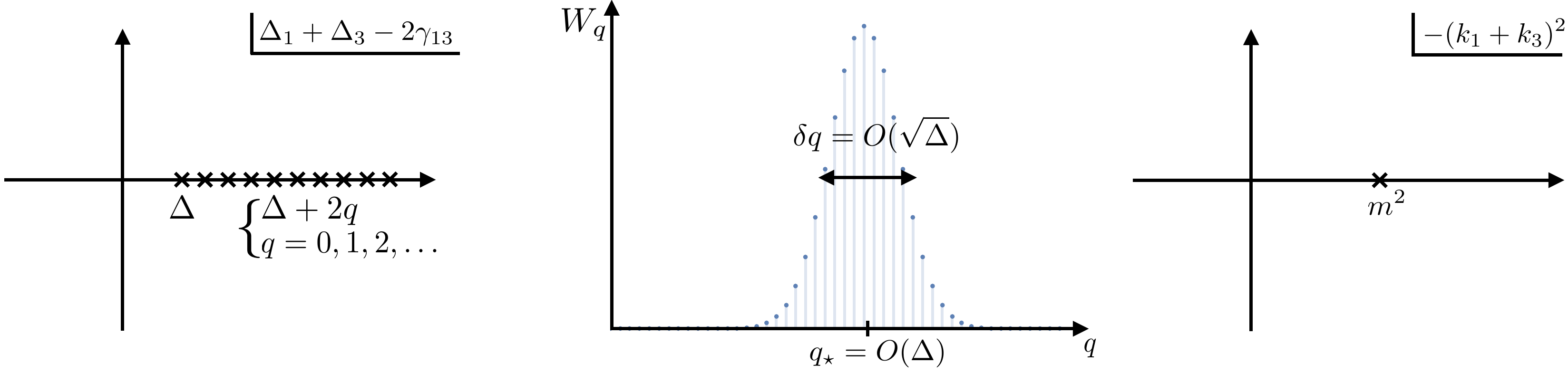}
\caption{\label{fig:FSLpoles}
The left figure shows the position of the poles of the Mellin amplitude associated with a tree-level scalar exchange Witten diagram. The middle figure shows the scaling of the residues of these poles in the flat space limit $\Delta \to \infty$. On the right, we show the resulting analytic structure for the scattering amplitude. The infinite sequence of poles of the Mellin amplitude gives rise to a single pole in the scattering amplitude.}
\end{center}
\end{figure}

Let us see how   formula (\ref{eq:FSLformula}) works for a scalar exchange diagram in
AdS. The associated Mellin amplitude was computed in \cite{Penedones:2010ue},
\[
M=-g^{2}R^{5-d}\mathcal{N}\sum_{q=0}^{\infty}\frac{W_{q}}{\Delta_{1}+\Delta_{3}-2\gamma_{13}-\Delta-2q}\ ,
\]
with
\[
W_{q}=\frac{\Gamma\left(\frac{\Delta_{1}+\Delta_{3}+\Delta-d}{2}\right)\Gamma\left(\frac{\Delta_{2}+\Delta_{4}+\Delta-d}{2}\right)}{2\Gamma\left(\frac{\sum_{i}\Delta_{i}-d}{2}\right)}\frac{\left(1+\frac{\Delta-\Delta_{1}-\Delta_{3}}{2}\right)_{q}\left(1+\frac{\Delta-\Delta_{2}-\Delta_{4}}{2}\right)_{q}}{q!\Gamma\left(\Delta-\frac{d}{2}+1+q\right)}
\]
In the limit of large $\Delta$'s (of the same order), the residues $W_q$
peak around $q=q_{\star}=O(\Delta)$ with a width $\delta q=O(\sqrt{\Delta})$.
More precisely,
\[
W_{q}=\frac{\exp\left[-\frac{(q-q_{\star})^{2}}{2\delta q^{2}}\right]}{\sqrt{2\pi}\left(\Delta_{1}+\Delta_{2}+\Delta_{3}+\Delta_{4}\right)\delta q}\left[1+O\left(\frac{1}{\sqrt{\Delta}}\right)\right]
\]
with
\begin{align}
q_{\star}&=\frac{\left(\Delta_{1}+\Delta_{3}-\Delta\right)\left(\Delta_{2}+\Delta_{4}-\Delta\right)}{2\left(\Delta_{1}+\Delta_{2}+\Delta_{3}+\Delta_{4}\right)}\ ,\\
\delta q^{2}&=\frac{\left(\Delta_{1}+\Delta_{3}+\Delta\right)\left(\Delta_{2}+\Delta_{4}+\Delta\right)\left(\Delta_{1}+\Delta_{3}-\Delta\right)\left(\Delta_{2}+\Delta_{4}-\Delta\right)}{2\left(\Delta_{1}+\Delta_{2}+\Delta_{3}+\Delta_{4}\right)^{3}}
\end{align}
 This  is depicted in figure \ref{fig:FSLpoles}.
We can then approximate the sum over $q$ by an integral, to obtain
\begin{eqnarray*}
\sum_{q=0}^{\infty}\frac{W_{q}}{\Delta_{1}+\Delta_{3}-2\gamma_{13}-\Delta-2q} & \approx & \int dq\frac{W_{q}}{\Delta_{1}+\Delta_{3}-2\gamma_{13}-\Delta-2q}\\
 & \approx & \frac{1}{\Delta_{1}+\Delta_{2}+\Delta_{3}+\Delta_{4}}\frac{1}{\Delta_{1}+\Delta_{3}-2\gamma_{13}-\Delta-2q_{\star}}\\
 & \to & -\frac{1}{R^{2}}\frac{1}{(k_{1}+k_{3})^{2}+m^{2}}
\end{eqnarray*}
Finally, we conclude that the flat space limit formula (\ref{eq:FSLformula}) leads to
\[
T=g^{2}\frac{1}{(k_{1}+k_{3})^{2}+m^{2}}
\]
as expected.

\subsection{Perturbative derivation} \la{derivationPer}

Now consider more general contact interactions involving derivatives
$g\nabla\dots\nabla\phi_{1}\dots\nabla\dots\nabla\phi_{n}$. 
In order
to determine the contact Witten diagram associated with this vertex we start
by computing the covariant derivative
\[
\nabla_{A}\frac{1}{(-2P\cdot X)^{\Delta}}=2\Delta\frac{P_{A}+(P\cdot X)X_{A}}{(-2P\cdot X)^{\Delta+1}}\equiv-\Delta\frac{Q_{A}}{(-2P\cdot X)^{\Delta}}
\]
where it is convenient to introduce the notation
\[
Q_A \colonequals \frac{P_{A}+(P\cdot X)X_{A}}{(P\cdot X)}
%\tilde{P}_{A}\equiv P_{A}+(P\cdot X)X_{A}=(P\cdot X)Q_{A}\ .
\]
Here we used the fact that covariant derivatives in AdS can be computed as partial derivatives in the embedding space projected to the tangent space of AdS
\cite{Costa:2016hju}. 
%, using the projector $\delta_A^B+X_AX^B$.
We have set $R=1$ to avoid cluttering the equations. Notice that
\[
\nabla_{B}(P_{A}+(P\cdot X)X_{A})=\left(\eta_{AB}+X_{A}X_{B}\right)(P\cdot X)\equiv(P\cdot X)G_{AB}
% \nabla_{B}\tilde{P}_{A}=\left(\eta_{AB}+X_{A}X_{B}\right)(P\cdot X)\equiv(P\cdot X)G_{AB}
\]
 and that $\nabla_{C}G_{AB}=0$ because $G_{AB}$ is the AdS metric.
By iterating these derivatives, we conclude that $l$ covariant derivatives lead to
\begin{eqnarray*}
\nabla_{A_{1}}\dots\nabla_{A_{l}}\frac{1}{(-2P\cdot X)^{\Delta}} 
%& = & \sum_{k=0}^{\left[\frac{l}{2}\right]}\frac{(-1)^{l-k}(\Delta)_{l-k}}{(-2P\cdot X)^{\Delta}}\frac{1}{(l-2k)!k!2^{k}}\sum_{perm}G_{A_{1}A_{2}}\dots G_{A_{2k-1}A_{2k}}Q_{A_{2k+1}}\dots Q_{A_{l}}\\
& = & \sum_{k=0}^{\left[\frac{l}{2}\right]}
\sum_{perm \atop \sigma} c_k(\sigma)\,
\frac{G_{A_{1}A_{2}}\dots G_{A_{2k-1}A_{2k}}Q_{A_{2k+1}}\dots Q_{A_{l}}}{(-2P\cdot X)^{\Delta}}
\end{eqnarray*}
for some coefficients $c_k(\sigma)$
where $\sigma$ labels the permutations of the indices $\{ A_1, \dots, A_l\}$.
Furthermore, it is not hard to see that $c_k\sim \Delta^{l-k}$ for large $\Delta$. Thus, the terms with $k=0$ dominate at large $\Delta$ and we find
\begin{eqnarray*}
\nabla_{A_{1}}\dots\nabla_{A_{l}}\frac{1}{(-2P\cdot X)^{\Delta}}
 & \approx & \frac{(-\Delta)^{l}}{(-2P\cdot X)^{\Delta}}Q_{A_{1}}\dots Q_{A_{l}}\,.
\end{eqnarray*}
This means that the correlation function is given by 
\[
\left\langle \mathcal{O}_{1}(P_{1})\dots\mathcal{O}_{n}(P_{n})\right\rangle \approx g\int_{AdS}dX\prod_{i=1}^{n}\frac{(-\Delta_{i})^{\alpha_{i}}\sqrt{\mathcal{C}_{\Delta_{i}}}}{(-2P_{i}\cdot X)^{\Delta_{i}}}\prod_{i<j}\left(Q_{i}\cdot Q_{j}\right)^{\alpha_{ij}}
\]
where $\alpha_{ij}$ are the number of contractions between derivatives
acting on $\phi_{i}$ and $\phi_{j}$ in the interaction vertex. We
also used $\alpha_{i}=\sum_{j}\alpha_{ij}$ for the total number of
derivatives acting on field $\phi_{i}$. The inner product
\[
Q_{i}\cdot Q_{j}=1-2\frac{(-2P_{i}\cdot P_{j})}{(-2P_{i}\cdot X)(-2P_{j}\cdot X)}
\]
gives rise to the same type of integrals as the pure contact diagram
studied above.
More precisely, we obtain a linear combination of terms of the form
\beq
D_{\Delta_1+\Lambda_1 \dots \Delta_n+\Lambda_n}(P_1, \dots, P_n)
\prod_{i<j}^n (-2P_i\cdot P_j)^{\lambda_{ij}}
\label{shiftedDfunctions}
\eeq
where $\lambda_{ij}$ are non-negative integers, $\Lambda_i=\sum_j \lambda_{ij}$ and 
\beq
D_{\Delta_1\dots \Delta_n}=\int_{AdS}dX\prod_{i=1}^n \frac{1}{(-2P_i\cdot X)^{\Delta_i}}\ .
\eeq
The Mellin amplitudes of (\ref{shiftedDfunctions}) are given by  \cite{Penedones:2010ue}
\beq
\frac{\left(\frac{1}{2}\sum_k \Delta_k -\frac{d}{2} \right)_{\sum_{i<j} \lambda_{ij}} }
{\prod_i (\Delta_i)_{\Lambda_i}}
 \prod_{i<j}(\gamma_{ij})_{\lambda_{ij}}
\eeq
times a constant independent of $\lambda_{ij}$. For large $\gamma_{ij}\sim \Delta_i$ we can approximate the Pochhammer symbols by powers
\beq
\frac{\left(\frac{1}{2}\sum_k \Delta_k -\frac{d}{2} \right)_{\sum_{i<j} \lambda_{ij}} }
{\prod_i (\Delta_i)_{\Lambda_i}}
 \prod_{i<j}(\gamma_{ij})_{\lambda_{ij}} \approx
 \prod_{i<j}\left(\frac{2\gamma_{ij}}{\Delta_i\Delta_j\sum_k \Delta_k}\right)^{\lambda_{ij}}\ .
\eeq
We conclude that, at large $\gamma_{ij}\sim \Delta_i$, 
 the Mellin amplitude can be obtained with the simple replacement rule
\[
Q_{i}\cdot Q_{j}\to1-\frac{\gamma_{ij}}{\Delta_{i}\Delta_{j}}\sum_{k=1}^{n}\Delta_{k}\ .
\]
This leads to 
\[
M\approx\frac{1}{2}gR^{n\frac{1-d}{2}+d+1-N}\pi^{\frac{d}{2}}\Gamma\left(\frac{\sum\Delta_{i}-d}{2}\right)\prod_{i=1}^{n}\frac{\sqrt{\mathcal{C}_{\Delta_{i}}}}{\Gamma(\Delta_{i})}\prod_{i<j}\left(\Delta_{i}\Delta_{j}-\gamma_{ij}\sum_{k=1}^{n}\Delta_{k}\right)^{\alpha_{ij}}
\]
where $N=2\sum_{i<j}\alpha_{ij}$ is the total number of derivatives
in the interaction vertex and we reintroduced the necessary factors of $R$ to make the expression dimensionless. Applying the flat space limit formula (\ref{eq:FSLformula})
we obtain the scattering amplitude
\[
T=g\prod_{i<j}\left(-k_{i}\cdot k_{j}\right)^{\alpha_{ij}}\ .
\]
We conclude that formula (\ref{eq:FSLformula})  works for any contact interaction with an arbitrary number of derivatives.
Since any diagram involving massive particles can be expanded as an infinite sum of contact interactions with derivatives (i.e. we can integrate out the massive particles) then this example provides a (perturbative) proof of formula (\ref{eq:FSLformula}).

\subsection{S-matrix analyticity and factorization from the OPE}

The Mellin amplitude has a simple analytic structure entirely controlled
by the  OPE of the conformal theory. In particular, if we
assume a generic discrete spectrum of scaling dimensions without degeneracies,
then the Mellin amplitude is meromorphic with simple poles at
\[
\gamma_{LR}\equiv\sum_{a\in L}\sum_{j\in R}\gamma_{aj}=\Delta-l+2q\ ,\qquad\qquad q=0,1,2,\dots
\]
where $L$ and $R$ are two disjoint sets whose union is $\{1, \dots, n\}$. 
$\Delta$ and $l$ are the dimension and spin of an operator
that appears in the OPEs $\prod_{a\in L}\mathcal{O}_{a}$ and $\prod_{j\in R}\mathcal{O}_{j}$.
Moreover, the residue of this pole is completely fixed by the OPE
coefficient (function) of this operator in these OPEs. In fact, one
can write factorization formulas for the residues in terms of lower
point Mellin amplitudes \cite{Fitzpatrick:2011ia,Paulos:2011ie}.

In order to reproduce the expected factorization pole in flat space,
we need that the sum over the satellite poles $q$ localizes around
\begin{equation}
q_{\star}=\frac{\left(\Delta-\sum_{a\in L}\Delta_{a}\right)\left(\Delta-\sum_{j\in R}\Delta_{j}\right)}{2\sum_{i}\Delta_{i}}\label{eq:qstarfac}
\end{equation}
This leads to
\[
\frac{1}{\gamma_{LR}-\Delta+l-2q_{\star}}\to\frac{\sum_{i}\Delta_{i}}{R^{2}}\frac{1}{k_{L}\cdot k_{R}-m^{2}}
\]
where $k_{L}=\sum_{a\in L}k_{a}$ is the momenta injected on the left
part of the amplitude and similarly for $k_{R}$. Notice that in the
flat space limit the spin $l$ is kept fixed while $\Delta\approx mR \to\infty$.

\subsubsection{Factorization on scalar particle}

Let us see how this works when the exchanged operator is a scalar.
In this case, the residues of the Mellin amplitude are given by \cite{Fitzpatrick:2011ia, Goncalves:2014rfa}
\[
\mathcal{Q}_{q}=\frac{-2\Gamma(\Delta)q!}{\left(\Delta-\frac{d}{2}+1\right)_{q}}L_{q}R_{q}\ ,\qquad\qquad L_{q}=\sum_{{n_{ab}\ge0\atop \sum n_{ab}=q}}M_{L}(\gamma_{ab}+n_{ab})\prod_{1\le a<b\le k}\frac{(\gamma_{ab})_{n_{ab}}}{n_{ab}!}
\]
and similarly for $R_{q}$. Here we are dividing the $n$ external
legs into a left group from 1 to $k$ and a right group from $k+1$
to $n$. We shall assume that the Mellin amplitudes $M_{L}$ and $M_{R}$
do not grow (or decay) exponentially for $\gamma_{ab}\sim\Delta_{a}\to\infty$.
On the other hand, 
\[
\prod_{1\le a<b\le k}\frac{(\gamma_{ab})_{n_{ab}}}{n_{ab}!}\approx e^{F_{L}}\prod_{1\le a<b\le k}\frac{1}{\sqrt{2\pi\delta n_{ab}^{2}}}\exp\left[-\frac{\left(n_{ab}-n_{ab}^{\star}\right)^{2}}{2\delta n_{ab}^{2}}\right]
\]
with
\begin{align}
F_{L}&=q\log\left(\frac{1+r_{L}}{r_{L}}\right)+\frac{q}{r_{L}}\log\left(1+r_{L}\right)\ ,  &r_{L}&=\frac{q}{\sum_{a<b}\gamma_{ab}}\ ,\\
n_{ab}^{\star}&=r_{L}\gamma_{ab}\ , &\delta n_{ab}^{2}&=r_{L}(1+r_{L})\gamma_{ab}\ .
\end{align}
This gives
\begin{eqnarray*}
L_{q} & \approx & M_{L}\left((1+r_{L})\gamma_{ab}\right)e^{F_{L}}\prod_{1\le a<b\le k}\int dn_{ab}\frac{1}{\sqrt{2\pi\delta n_{ab}^{2}}}\exp\left[-\frac{\left(n_{ab}-n_{ab}^{\star}\right)^{2}}{2\delta n_{ab}^{2}}\right]\delta\left(q-\sum n_{ab}\right)\\
 & = & M_{L}\left((1+r_{L})\gamma_{ab}\right)e^{F_{L}}\int\frac{ds}{2\pi}\prod_{1\le a<b\le k}\int dn_{ab}\frac{\exp\left[-\frac{\left(n_{ab}-n_{ab}^{\star}\right)^{2}}{2\delta n_{ab}^{2}}+is\left(n_{ab}-n_{ab}^{\star}\right)\right]}{\sqrt{2\pi\delta n_{ab}^{2}}}\\
 & = & M_{L}\left((1+r_{L})\gamma_{ab}\right)e^{F_{L}}\frac{1}{\sqrt{2\pi q(1+r_{L})}}
\end{eqnarray*}
Putting things together we find
\[
\mathcal{Q}_{q}\approx-2\frac{\left(\Delta+q\right)^{\frac{d-1}{2}}\exp\left[F_{0}+F_{L}+F_{R}\right]}{\Delta^{\frac{d}{2}}\sqrt{q(1+r_{L})(1+r_{R})}}M_{L}\left((1+r_{L})\gamma_{ab}\right)M_{R}\left((1+r_{R})\gamma_{ij}\right)
\]
where 
\[
F_{0}=2\Delta\log\Delta-\Delta+q\log q-(q+\Delta)\log(q+\Delta)\ .
\]
The sum over $q$ is also dominated by a saddle point,
\[
F_{0}+F_{L}+F_{R}\approx F_{\star}-\frac{(q-q_{\star})^{2}}{2\delta q^{2}}
\]
where $q_{\star}$ is given by (\ref{eq:qstarfac}) and 
\[
F_{\star}=2\Delta\log\Delta-\Delta+\frac{\Sigma}{2}\log\frac{\Sigma}{2}-\frac{\Sigma_{L}}{2}\log\frac{\Sigma_{L}}{2}-\frac{\Sigma_{R}}{2}\log\frac{\Sigma_{R}}{2}
\]
 
\[
\delta q^{2}=\frac{q_{\star}}{(1+r_{L})(1+r_{R})}=\frac{\left(\Sigma-\Sigma_{L}\right)\left(\Sigma-\Sigma_{R}\right)\Sigma_{L}\Sigma_{R}}{2\Sigma^{3}}
\]
with
\[
\Sigma=\sum_{i=1}^{n}\Delta_{i}\ ,\qquad\Sigma_{L}=\Delta+\sum_{a\in L}\Delta_{a}\ ,\qquad\Sigma_{R}=\Delta+\sum_{j\in R}\Delta_{j}\ .
\]
The contribution from the poles of the Mellin amplitude is given by
\begin{eqnarray*}
&&\sum_{q=0}^\infty \frac{Q_q}{\gamma_{LR}-\Delta-2q} \approx
\int_0^\infty dq \frac{Q_q}{\gamma_{LR}-\Delta-2q}
\\ & \to & -2\frac{\left(\Delta+q_{\star}\right)^{\frac{d-1}{2}}\exp\left[F_{\star}\right]\sqrt{2\pi\delta q^{2}}}{\Delta^{\frac{d}{2}}\sqrt{q_{\star}(1+r_{L})(1+r_{R})}}M_{L}\left( \gamma_{ab}^L\right)M_{R}\left( \gamma_{ij}^R\right)\frac{\sum_{i}\Delta_{i}}{R^{2}}\frac{1}{k_{L}\cdot k_{R}-m^{2}}\\
 & = & 4\frac{\sqrt{2\pi}\exp\left[F_{\star}\right]}{R^{2}}\left[\frac{\Sigma_{L}\Sigma_{R}}{2\Sigma}\right]^{\frac{d+1}{2}}\Delta^{-\frac{d}{2}}M_{L}\left(\gamma_{ab}^{L}\right)M_{R}\left(\gamma_{ij}^{R}\right)\frac{1}{k_{L}^{2}+m^{2}}\\
 & \approx & \frac{1}{R^{2}}\frac{\mathcal{N}}{\mathcal{N}_{L}\mathcal{N}_{R}}M_{L}\left(\gamma_{ab}^{L}\right)M_{R}\left(\gamma_{ij}^{R}\right)\frac{1}{k_{L}^{2}+m^{2}}
\end{eqnarray*}
where
\[
\gamma_{ab}^{L}=(1+r_{L})\gamma_{ab}\to \frac{\Delta_{a}\Delta_{b}+R^{2}k_{a}\cdot k_{b}}{\Sigma_L}
\]
is exactly what it should be to correspond to the flat space limit of the left
Mellin amplitude. This leads to the factorization formula
\[
T\approx\frac{T_{L}T_{R}}{k_{L}^{2}+m^{2}}\,,
\]
with
\[
T_{L}=\lim_{\Delta\to\infty}\frac{1}{\mathcal{N}_{L}}R^{(k+1)\frac{d-1}{2}-d-1}M_{L}\left(\gamma_{ab}=\frac{\Delta_{a}\Delta_{b}+R^{2}k_{a}\cdot k_{b}}{\Sigma_{L}}\right)
\]
and similarly for $T_R$.

\subsubsection{Factorization of four-particle amplitude \label{sec:Fac4pt}}

The four point Mellin amplitude has a semi-infinite sequence of poles
associated to each primary operator exchanged,
\[
M\approx\frac{\lambda_{12k}\lambda_{34k}\mathcal{Q}_{l,q}(\gamma_{13})}{\gamma_{LR}-\Delta+l-2q}\ ,\qquad q=0,1,2,\dots
\]
where $\Delta$ and $l$ are the dimension and spin of the exchanged
operator $\mathcal{O}_{k}$ and $\lambda$'s are OPE coefficients. The residue
is a Mack polynomial of degree $l$ in the Mellin variable $\gamma_{13}$.
We follow the conventions of \cite{Costa:2012cb},
\begin{eqnarray}
\mathcal{Q}_{l,q}(\gamma_{13}) & = & -\frac{2\Gamma(\Delta+l)(\Delta-1)_{l}}{4^{l}\Gamma\left(\frac{\Delta+l+\Delta_{12}}{2}\right)\Gamma\left(\frac{\Delta+l-\Delta_{12}}{2}\right)\Gamma\left(\frac{\Delta+l+\Delta_{34}}{2}\right)\Gamma\left(\frac{\Delta+l-\Delta_{34}}{2}\right)}
\label{Mackpol}\\
 &  & \frac{Q_{l,q}(\gamma_{13})}{q!\left(\Delta-\frac{d}{2}+1\right)_{q}\Gamma\left(\frac{\Delta_{1}+\Delta_{2}-\Delta+l}{2}-q\right)\Gamma\left(\frac{\Delta_{3}+\Delta_{4}-\Delta+l}{2}-q\right)}\,.
 \nonumber
\end{eqnarray}
In the flat space limit $\gamma_{13}\sim\Delta\sim q\sim\Delta_{i}\gg1$
we find
\begin{align}
Q_{l,q}(\gamma_{13})\approx&
\frac{l!}{2^l\left(\frac{d}{2}-1\right)_l } \Delta^{-2l}
\left[\left(\Delta^{2}-\Delta_{12}^{2}\right)\left(\Delta^{2}-\Delta_{34}^{2}\right)q(\Delta+q)\right]^{\frac{l}{2}}
\\
&C_{l}\left(\frac{2q\left(\Delta^{2}+\Delta_{12}\Delta_{34}\right)+\Delta\left[\left(\Delta+\Delta_{12}\right)\left(\Delta+\Delta_{34}\right)-4\gamma_{13}\Delta\right]}{2\sqrt{q(q+\Delta)\left(\Delta^{2}-\Delta_{12}^{2}\right)\left(\Delta^{2}-\Delta_{34}^{2}\right)}}\right)\,,
\nonumber
\end{align}
where $C_{l}(z)\equiv C_{l}^{\left(\frac{d-2}{2}\right)}(z)$ is the Gegenbauer polynomial appropriate for spin $l$
partial waves in $(d+1)$-dimensional flat spacetime. The $q$ dependence
is polynomial. In the flat space limit, the other factors present  in (\ref{Mackpol}) are peaked around $q=q_\star$ with
\[
q_{\star}=\frac{\left(\Delta_{1}+\Delta_{3}-\Delta\right)\left(\Delta_{2}+\Delta_{4}-\Delta\right)}{2\left(\Delta_{1}+\Delta_{2}+\Delta_{3}+\Delta_{4}\right)}\ .
\]
This together with the flat space limit rule $\gamma_{13} \to (\Delta_1\Delta_3+R^2 k_1\cdot k_3)/(\sum_i \Delta_i)$ simplifies the argument of the Gegenbauer polynomial to 
\begin{align}
\cos\theta=\frac{\left(s+m_{1}^{2}-m_{2}^{2}\right)\left(s+m_{3}^{2}-m_{4}^{2}\right)-4s\,k_{1}\cdot k_{3}}{4\sqrt{\left[(k_{1}\cdot k_{2})^{2}-m_{1}^{2}m_{2}^{2}\right]\left[(k_{3}\cdot k_{4})^{2}-m_{3}^{2}m_{4}^{2}\right]}}\,,
\end{align}
where $\theta$ is the usual scattering angle and $s=-(k_{1}+k_{2})^{2}$. This gives the following factorization
formula for scattering amplitudes
\[
T\approx\frac{T_{L}T_{R}}{(k_{1}+k_{2})^{2}+m^{2}} 
\frac{l!\,C_{l}\left(\cos\theta\right)}{2^l\left(\frac{d}{2}-1\right)_l} 
\label{eq:FacT4pt}
\]
with
\begin{eqnarray*}
T_{L} 
 & = & \lim_{\Delta\to\infty}\frac{1}{\mathcal{N}}R^{\frac{d-5}{2}}\frac{\lambda_{12k}}{\Gamma\left(\frac{\Delta_{1}+\Delta_{2}-\Delta}{2}\right)\Gamma\left(\frac{\Delta+\Delta_{2}-\Delta_{1}}{2}\right)\Gamma\left(\frac{\Delta+\Delta_{1}-\Delta_{2}}{2}\right)}
\end{eqnarray*}
where $\mathcal{N}$ is given by
\begin{align}
\mathcal{N}&=\frac{1}{2}\pi^{\frac{d}{2}}\Gamma\left(\frac{\Delta_{1}+\Delta_{2}+\Delta-d}{2}\right)\frac{\sqrt{\mathcal{C}_{\Delta_{1}}}}{\Gamma(\Delta_{1})}\frac{\sqrt{\mathcal{C}_{\Delta_{2}}}}{\Gamma(\Delta_{2})}\frac{\sqrt{\mathcal{C}_{\Delta}}}{\Gamma(\Delta)}\ .
\end{align}

\subsubsection{Two-particle cut} \la{Two-particleSec}

Let us consider a one-loop diagram that gives rise to a two-particle
cut in the scattering amplitude
\begin{align}
T&=g^{2}\int\frac{d^{D}q}{\left(2\pi\right)^{D}}\frac{1}{q^{2}+m^{2}}\frac{1}{\left(q-p\right)^{2}+\bar{m}^{2}}\\
&=g^{2}\frac{\Gamma\left(2-\frac{D}{2}\right)}{\left(4\pi\right)^{\frac{D}{2}}}\int_{0}^{1}dt\left[t(1-t)p^{2}+t\,m^{2}+(1-t)\bar{m}^{2}\right]^{\frac{D}{2}-2}
\end{align}
We are assuming $D<4$ in order to get a UV finite integral. This
amplitude can also be written in the Kallen-Lehmann spectral representation
\[
T=\int_{\left(m+\bar{m}\right)^{2}}^{\infty}d\mu^{2}\frac{\rho(\mu^{2})}{p^{2}+\mu^{2}}
\]
where the spectral density of the two-particle cut is given by 
\begin{eqnarray*}
\rho(\mu^{2}) & = & \frac{T(p^{2}=-\mu^{2}-i\epsilon)-T(p^{2}=-\mu^{2}+i\epsilon)}{2\pi i}\\
 & = & g^{2}\frac{\Gamma\left(2-\frac{D}{2}\right)}{\left(4\pi\right)^{\frac{D}{2}}}\int_{0}^{1}\frac{dt}{2\pi i}\left[\left(-z(t,\mu)-i\epsilon\right)^{\frac{D}{2}-2}-\left(-z(t,\mu)+i\epsilon\right)^{\frac{D}{2}-2}\right]\\
 & = & -g^{2}\frac{1}{\pi}\frac{\Gamma\left(2-\frac{D}{2}\right)}{\left(4\pi\right)^{\frac{D}{2}}}\sin\frac{\pi D}{2}\int_{0}^{1}dt\,\Theta\left(z(t,\mu)\right)\left(z(t,\mu)\right)^{\frac{D}{2}-2}\\
 & = & -g^{2}\frac{1}{\pi}\frac{\Gamma\left(2-\frac{D}{2}\right)}{\left(4\pi\right)^{\frac{D}{2}}}\sin\frac{\pi D}{2}\,\mu^{D-4}\int_{t_{1}}^{t_{2}}dt\,\left((t-t_{1})(t_{2}-t)\right)^{\frac{D}{2}-2}\\
 & = & -g^{2}\frac{1}{\pi}\frac{\Gamma\left(2-\frac{D}{2}\right)}{\left(4\pi\right)^{\frac{D}{2}}}\sin\frac{\pi D}{2}\,\mu^{D-4}(t_{2}-t_{1})^{D-3}\frac{\Gamma^{2}\left(\frac{D}{2}-1\right)}{\Gamma\left(D-2\right)}\\
 & = & g^{2}\frac{4^{2-D}}{2\pi^{\frac{D-1}{2}}\Gamma\left(\frac{D-1}{2}\right)}\mu^{2-D}\left[\mu^{4}+m^{4}+\bar{m}^{4}-2\mu^{2}(m^{2}+\bar{m}^{2})-2m^{2}\bar{m}^{2}\right]^{\frac{D-3}{2}}
\end{eqnarray*}
where $z(t,\mu)=t(1-t)\mu^{2}-t\,m^{2}-(1-t)\bar{m}^{2}=\mu^{2}(t-t_{1})(t_{2}-t_{1})$.
It is convenient to write $\mu=m+\bar{m}+2y$ to find
\begin{equation}
T=\int_{0}^{\infty}dy\frac{\tilde{\rho}(y)}{p^{2}+\left(m+\bar{m}+2y\right)^{2}}\,,
\label{eq:SpectralRepy}
\end{equation}
with
\[
\tilde{\rho}(y)=g^{2}\frac{1}{2\pi^{\frac{D-1}{2}}\Gamma\left(\frac{D-1}{2}\right)}\left(m+\bar{m}+2y\right)^{3-D}\left[y\left(m+y\right)\left(\bar{m}+y\right)\left(m+\bar{m}+y\right)\right]^{\frac{D-3}{2}}\,.
\]
The corresponding loop diagram in AdS can be computed with the use
of the following identity \cite{Fitzpatrick:2011hu}
\[
G_{\Delta}(X,Y)G_{\bar{\Delta}}(X,Y)=\sum_{n=0}^{\infty}a_{n}G_{\Delta+\bar{\Delta}+2n}(X,Y)
\]
where
\[
a_{n}=\frac{1}{R^{D-2}}\frac{\left(\frac{D-1}{2}\right)_{n}\left(\Delta+\bar{\Delta}+2n\right)_{\frac{3-D}{2}}\left(\Delta+\bar{\Delta}+n+2-D\right)_{n}}{2\pi^{\frac{D-1}{2}}n!\left(\Delta+n\right)_{\frac{3-D}{2}}\left(\bar{\Delta}+n\right)_{\frac{3-D}{2}}\left(\Delta+\bar{\Delta}+n-\frac{D-1}{2}\right)_{n}}
\]
This means that the loop is equivalent to an infinite sum of scalar
exchanges like the ones we studied in section \ref{sub:Scalar-exchange}.
We conclude that the flat space limit leads to 
\[
T=\lim_{R\to\infty}\sum_{n=0}^{\infty}a_{n}\frac{g^{2}}{p^{2}+\left(\Delta+\bar{\Delta}+2n\right)^{2}/R^{2}}
\]
where the limit should be taken with $m=\Delta/R$ and $\bar{m}=\bar{\Delta}/R$
fixed. Remarkably, in this limit, we find
\[
a_{n}\approx\frac{1}{R}\tilde{\rho}(y)
\]
where we also kept $y=n/R$ fixed. This leads directly to the spectral
representation (\ref{eq:SpectralRepy}).

\subsection{Many point functions}

When $n\ge d+3$ not all $\frac{1}{2}n(n-3)$ cross ratios are independent. This follows from the fact that $d+3$ vectors $P_i \in \mathbb{R}^{d+1,1}$ can not be linearly independent.
Therefore, the determinant of the $(d+3)\times(d+3)$ matrix with entries $(-2P_i \cdot P_j )$ vanishes,
\beq
\det_{i,j} \,(-2P_i \cdot P_j ) =0\,.
\eeq 
This leads to non-uniqueness of the Mellin amplitude. The Mellin amplitude $M_0$ defined by
\beq
\Lambda(P_1,\dots,P_n) \det_{i,j} \,(-2P_i \cdot P_j )=
\int[d\gamma]M_0(\gamma_{ij})\prod_{1\le i<j\le n}\frac{\Gamma(\gamma_{ij})}{\left(-2P_{i}\cdot P_j\right)^{\gamma_{ij}}}\,,
\eeq
where $\Lambda$ is any Lorentz invariant homogeneous function (with appropriate weights), is equivalent to zero. In the flat space limit (large $\gamma_{ij}$),  we obtain
\beq
M_0(\gamma_{ij}) \approx M_\Lambda(\gamma_{ij}) \det_{i,j} \,(\gamma_{ij} )\ ,
\eeq
assuming that the Mellin amplitude $M_\Lambda$ does not depend exponentially on $\gamma_{ij}$ for large $\gamma_{ij}$.
Let us see what this gives under the flat space limit formula  (\ref{eq:FSLformula}),
\beq
 \det_{i,j} \,(\gamma_{ij} )\sim \det_{i,j} \,(m_im_j+ k_i \cdot k_j ) = 
 \det_{i,j} \,(  k_i \cdot k_j ) +\sum_{l=1}^{d+3} \det_{i,j} \,(\delta_{il} m_im_j+ k_i \cdot k_j )\,,
\eeq
where we have used linearity of the determinant with respect to each line of the matrix.
By expanding these determinants along line $l$, it is clear that all these determinants of  
$(d+3)\times(d+3)$ matrices can be written as linear combinations of determinants 
$\det_{i,j} \,(  k_i \cdot k_j )$ of  $(d+2)\times(d+2)$ matrices. But these must vanish because the momenta $k_i$ are $(d+1)$-dimensional vectors.
Therefore, the non-uniqueness of the Mellin amplitude as a function of $\gamma_{ij}$ maps into the same type of non-uniqueness of the scattering amplitudes when written in terms of the Mandelstam invariants.

\subsection{Relation to phase shift formula \label{sec:Mellintophaseshift}}

In this section we  show that the imaginary part of the scattering amplitude $T(s,t)$ obtained from the flat space limit formula \eqref{eq:FSLgeneralformula} using Mellin amplitudes agrees with the result that follows from the phase shift formula \eqref{eq:phaseshiftformula}. For simplicity, we restrict ourselves to the case of equal external operators and denote them by $\mathcal{O}_1$.

We saw in \ref{sec:Fac4pt} that each operator exchanged in the OPE gives rise to a pole in the scattering amplitude, whose residue is related to the product of OPE coefficients. Taking the imaginary part of equation \eqref{eq:FacT4pt}, we conclude that
\be
{\rm Im}\, T(s,t)  = \lim_{\Delta_1\to \infty} \sum_{\Delta,l}   W\,
 \delta(s-m^2)  C_{l}(z)
 \ ,\qquad\qquad z=\cos\theta=\frac{u-t}{u+t}\ ,
\label{eq:ImT}
\ee
where the sum runs over all primary operators $\mathcal{O}_{\Delta,l}$ with (even) spin $l$ and dimension $\Delta$ that appear in the OPE $\mathcal{O}_1\times \mathcal{O}_1$. The mass $m$ is given by $m = \lim_{\Delta_1\to \infty} \Delta \frac{m_1}{\Delta_1}$ and the weight $W$ is given by
\be
W%= \pi\,  T_L T_R \frac{2^l\left(\frac{d}{2}-1\right)_l}{l!}
= \pi\,
m_1^{5-d}  \Delta_1^{d-5}
\frac{l!\,}{2^l  \left(\frac{d}{2}-1\right)_l }
\frac{4
\Gamma^2\left(\Delta\right)
\Gamma^4\left(\Delta_{1}\right)
}{\pi^{d}
\Gamma^2\left(\frac{2\Delta_{1} +\Delta-d}{2}\right)
\Gamma^2\left(\frac{2\Delta_{1} -\Delta}{2}\right)
\Gamma^4\left(\frac{\Delta}{2}\right)
\mathcal{C}_{\Delta_1}^2
\mathcal{C}_\Delta}
\lambda_{\Delta,l}^2\ .
   \label{w1}
\ee

The 2 to 2 scattering amplitude of identical scalar particles in $(d+1)$ spacetime dimensions can also be written as
\[
T(s,t)=i\frac{2\sqrt{s}}{\left(s-4m^{2}\right)^{\frac{d-2}{2}}}\sum_{{l=0\atop even}}^{\infty}\left(1-e^{2i\delta_{l}(s)}\right)P_{l}^{(d)}(z)
\label{Tfromphaseshift}
\]
where $\delta_{l}(s)$ is the phase shift and 
\begin{align}
P_{l}^{(d)}(z)
&=2^{2 d-3} \pi ^{\frac{d}{2}-1}
   (d+2 l-2) \Gamma
   \left(\frac{d}{2}-1\right)
   C_l(z)
\end{align}
are harmonic polynomials on the sphere $S^{d-1}$ at spatial infinity.
The normalization of the polynomials was chosen in order to describe easily free propagation. More precisely, they lead to the following identity 
\begin{eqnarray*}
 &  & i(2\pi)^{d+1}\delta^{(d+1)}\left(\sum k_{i}\right)i\frac{2\sqrt{s}}{\left(s-4m^{2}\right)^{\frac{d-2}{2}}}\sum_{{l=0\atop even}}^{\infty}P_{l}^{(d)}(z)+\\
 &  & +4E_{1}E_{2}(2\pi)^{2d}\left[\delta^{(d)}(k_{1}-k_{3})\delta^{(d)}(k_{2}-k_{4})+\delta^{(d)}(k_{1}-k_{4})\delta^{(d)}(k_{2}-k_{3})\right]=0\ .
\end{eqnarray*} 

\begin{figure}
\begin{center}
\includegraphics[width=10cm]{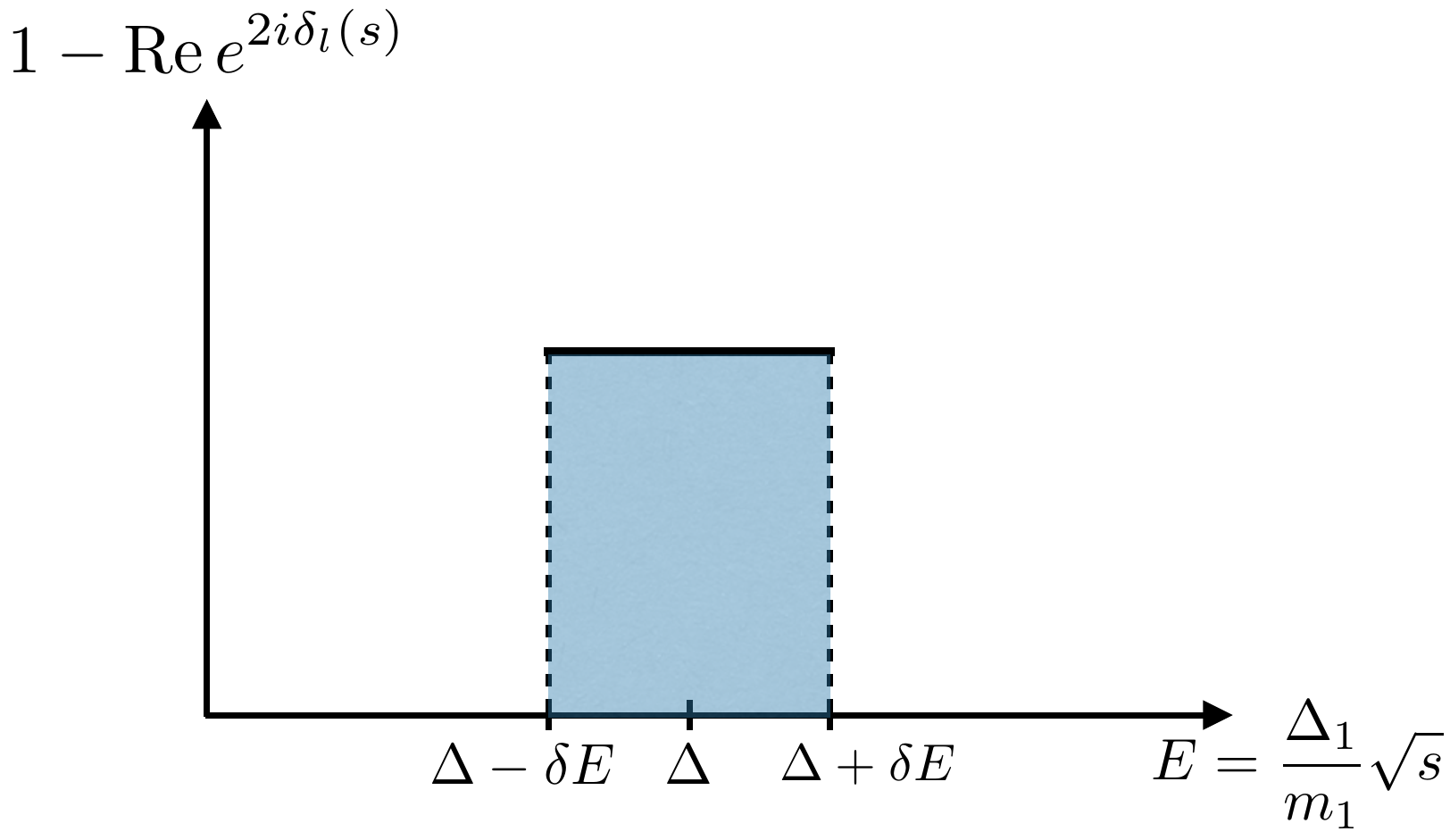}
\caption{\label{fig:Regularizeddelta}
Each operator of dimension $\Delta$ and spin $l$ makes a localized contribution to $1-{\rm Re}\, e^{2i\delta_{l}(s)}$. The height of the rectangle is given by \eqref{eq:heighofregtangle}.
In the flat space limit this contribution becomes proportional to $\delta(s-m^2)$. The sum over the contribution of all operators produces a smooth function of $s$.
}
\end{center}
\end{figure}

Taking the imaginary part of \eqref{Tfromphaseshift} we obtain
\be
{\rm Im}\, T(s,t)=\frac{2\sqrt{s}}{\left(s-4m_1^{2}\right)^{\frac{d-2}{2}}}\sum_{{l=0\atop even}}^{\infty}\left[1-{\rm Re}\, e^{2i\delta_{l}(s)} \right]P_{l}^{(d)}(z)\ .
\ee
From the phase shift flat space limit formula \eqref{eq:phaseshiftformula} we find
\[
1-{\rm Re}\, e^{2i\delta_{l}(s)}=\lim_{\Delta_1\to \infty} \frac{1}{N_{l}(E)}\sum_{|\Delta-E|<\delta E}
\left[w(\Delta)\lambda_{\Delta,l}\right]^{2}\left[1-\cos \pi(\Delta-2\Delta_{1}-l) \right]
\label{1minuscosdelta}
\]
This means that each operator of dimension $\Delta$ contributes a regularized delta-function to (\ref{1minuscosdelta}) as depicted in figure \ref{fig:Regularizeddelta}. 
More precisely, it contributes
\[
1-{\rm Re}\, e^{2i\delta_{l}(s)}=\lim_{\Delta_1 \to \infty} \frac{
\left[w(\Delta)\lambda_{\Delta,l}\right]^{2}\left[1-\cos \pi(\Delta-2\Delta_{1}-l) \right]}{N_{l}(\Delta)} 
\label{eq:heighofregtangle}
\]
if 
\be
\frac{m_1^2}{\Delta_1^2}(\Delta-\delta E)^2<s<\frac{m_1^2}{\Delta_1^2}(\Delta+\delta E)^2
\ee
and zero for other values of $s$. Thus,
\[
1-{\rm Re}\, e^{2i\delta_{l}(s)}= \lim_{\Delta_1 \to \infty}  \sum_{\Delta} \delta(s-m^2) 
\frac{4\delta E \Delta m_1^2}{\Delta_1^2}
\frac{\left[w(\Delta)\lambda_{\Delta,l}\right]^{2}\left[1-\cos \pi(\Delta-2\Delta_{1}-l) \right]}{N_{l}(\Delta)}\,, 
\]
which  leads to (\ref{eq:ImT}) with the following expression for the weight
\be
W=\frac{m\,
2^{2 d-2} \pi ^{\frac{d}{2}-1}
   (d+2 l-2) \Gamma
   \left(\frac{d}{2}-1\right) 
}{ \left(m^2-4m_1^{2}\right)^{\frac{d-2}{2}}} 
\frac{4\delta E\Delta m_1^2}{\Delta_1^2}
\frac{ \left[1-\cos \pi(\Delta-2\Delta_{1}-l) \right]}{N_{l}(\Delta)} 
\left[w(\Delta)\lambda_{\Delta,l}\right]^{2}\,.
\label{w2}
\ee
One can easily check that \eqref{w1} and \eqref{w2} are equivalent if
\be
N_l(\Delta)\approx \delta E \frac{\Delta ^{d/2} \left(\Delta
   -2 \Delta
   _1\right)^{-\frac{d}{2}  } \Delta
   _1^{d-4 \Delta _1}
   \left(\Delta +2 \Delta
   _1\right)^{-\frac{3
   d}{2}  }
   2^{3 d-4 \Delta _1+l  } \Gamma
   \left(\frac{d}{2}+l\right)}{\pi \, l!}    
   \label{asymNlDelta}
\ee
in the flat space limit  $\Delta\sim \Delta_1 \gg \delta E \gg l\sim1$. This asymptotic behaviour of the spectral weight
\be
N_l(E)= \sum_{|\Delta-E|<\delta E} \left[w(\Delta)\lambda_{\Delta,l}\right]^{2}
\ee
is the same as for generalized free fields. More precisely, one can check that the exact formula for free fields in AdS \cite{Fitzpatrick:2011dm}
\be
\lambda^2_{\Delta,l}=\frac{2^{l+1}
   \left[\left(\Delta _1-\frac{d}{2}+1\right)_n \left(\Delta_1\right)_{l+n}\right]^2}{l!
   n!
   \left(\frac{d}{2}+l\right)_n
   \left(n+2 \Delta_1-d+1\right)_n \left(l+2 n+2
   \Delta _1-1\right)_l
   \left(l+n+2
   \Delta _1-\frac{d}{2}\right)_n}\,,
\ee
with $\Delta=2\Delta_1+2n+l$, has exactly the asymptotic behavior \eqref{asymNlDelta}. 
This asymptotic behavior is also compatible with the general results of \cite{Pappadopulo:2012jk} but it is stronger.
We claim that QFT in AdS leads to this universal asymptotic form of the spectral density $N_l(\Delta)$.
This follows from the bulk wave-function construction of appendix \ref{sec:scatteringstates}.
The point is that this is the spectral density of a two particle state in AdS, where the particles are very well separated in the initial time slice. Therefore, locality of the interactions allows us to measure the energy distribution of the state reliably in the initial time slice.
This proves that the phase shift and the Mellin formulas lead to the same imaginary part of the flat space scattering amplitude.

\section{Large $\Delta$ limit of the crossing equations in $d=1$}
\label{appendix:largedelta}
We are interested in studying bounds on the dimension $\Delta_2$ of the leading scalar in the OPE of some other scalar $\mathcal{O}_1$ with itself, when the dimension $\Delta_1$ of the latter is very large. More precisely, we take \emph{all} the scaling dimensions of nontrivial operators to be large. We will show that optimal bounds require a number of derivatives that is at least as large as $\sqrt{\Delta_1}$. We begin by considering the large dimension limit of $d=1$ conformal blocks. This is remarkably simple:
\bea
G_\Delta(z)=\frac{(4 \rho)^{\Delta}}{\sqrt{1-\rho^2}}
\left[ 1+O\left(\frac{1}{\Delta}\right)\right]\,, \qquad \rho=\frac{z}{(1+\sqrt{1-z})^2}\,.
\eea
The statement of crossing symmetry of a four point function is
\bea
\sum \lambda_{\Delta}^2 F_{\Delta}(z)\equiv \sum \lambda^2_{\Delta} \left( (1-z)^{2\Delta_1} G_{\Delta}(z)-z^{2\Delta_1} G_{\Delta}(1-z)\right)=0\,.
\eea
As usual, we try to rule out solutions to these constraints by constructing a linear functional with certain positivity properties. We shall take the functional to be a sum of derivatives with respect to $z$ at $z=1/2$. Then it can be shown that
\bea
\partial_z^n F_{\Delta}(z)\bigg|_{z=1/2}=\frac{(4\rho)^\Delta}{\sqrt{1-\rho^2}} \left(-\frac{2\,x_\Delta^{n}}{4^{\Delta_1-n}}\right), \quad x_\Delta\equiv \Delta_1-\frac{\sqrt{2}}2\,\Delta, \qquad \rho=3-2\sqrt{2},
\eea
for $\Delta,\Delta_1\gg 1$ and $n$ odd (zero otherwise). This approximation captures only the leading term. We also need the behaviour of the identity block, for which $\Delta=0$. We have
\bea
\partial_z^n F_{0}(z)\bigg|_{z=1/2}=-\frac{2 \Delta_1^n}{4^{\Delta_1-n}} , \qquad \mbox{for odd $n$}
\eea
We can now find very simple solutions to crossing symmetry independent of the number of derivatives. The contribution of the identity can be cancelled by a vector with 
\bea
x_\Delta=-\Delta_1 \Leftrightarrow \Delta =2\sqrt{2} \Delta_1
\eea
Similarly, the contribution of any vector with $\Delta_2<\sqrt{2} \Delta_1$ may be cancelled by one with $\Delta=2\sqrt{2} \Delta_1-\Delta_2$. This is a special case of the approximate reflection symmetry discussed in \cite{Kim:2015oca}. In particular the solution exists for any number of derivatives, as long as we take the $\Delta_1\to \infty$ limit first. Since such solutions exist, whatever bounds one finds can never rule it out. More precisely, for finite $\Delta_1$ we expect that the extremal solution will be given by sets of vectors which closely cluster around the peaks determined by these equations, and this is indeed borne out by explicit numerical checks. 

In the above we approximated the $n$-th derivative of factors such as $z^\Delta$ by $\Delta^n z^{\Delta-n}$. In reality one obtains Pochhammer symbols, which for $\Delta\gg n\gg 1$ become
\bea
(\Delta)_n\simeq \Delta^n \left(1+\frac {n^2}{2\Delta}+\ldots\right).
\eea
This shows that corrections to the results derived above will only kick in if $n$ is at least $O(\sqrt{\Delta})$.

\bibliography{Biblio}% Produces the bibliography via BibTeX.

\providecommand{\href}[2]{#2}\begingroup\raggedright\begin{thebibliography}{10}

\bibitem{Ferrara:1973yt}
S.~Ferrara, A.~F. Grillo, and R.~Gatto, {\it {Tensor representations of
  conformal algebra and conformally covariant operator product expansion}},
  {\em Annals Phys.} {\bf 76} (1973) 161--188.

\bibitem{Polyakov:1974gs}
A.~M. Polyakov, {\it {Nonhamiltonian approach to conformal quantum field
  theory}},  {\em Zh. Eksp. Teor. Fiz.} {\bf 66} (1974) 23--42.

\bibitem{Rattazzi:2008pe}
R.~Rattazzi, V.~S. Rychkov, E.~Tonni, and A.~Vichi, {\it {Bounding scalar
  operator dimensions in 4D CFT}},  {\em JHEP} {\bf 12} (2008) 031,
  [\href{http://arxiv.org/abs/0807.0004}{{\tt arXiv:0807.0004}}].

\bibitem{Rychkov:2009ij}
V.~S. Rychkov and A.~Vichi, {\it {Universal Constraints on Conformal Operator
  Dimensions}},  {\em Phys. Rev.} {\bf D80} (2009) 045006,
  [\href{http://arxiv.org/abs/0905.2211}{{\tt arXiv:0905.2211}}].

\bibitem{Caracciolo:2009bx}
F.~Caracciolo and V.~S. Rychkov, {\it {Rigorous Limits on the Interaction
  Strength in Quantum Field Theory}},  {\em Phys. Rev.} {\bf D81} (2010)
  085037, [\href{http://arxiv.org/abs/0912.2726}{{\tt arXiv:0912.2726}}].

\bibitem{ElShowk:2012ht}
S.~El-Showk, M.~F. Paulos, D.~Poland, S.~Rychkov, D.~Simmons-Duffin, and
  A.~Vichi, {\it {Solving the 3D Ising Model with the Conformal Bootstrap}},
  {\em Phys.Rev.} {\bf D86} (2012) 025022,
  [\href{http://arxiv.org/abs/1203.6064}{{\tt arXiv:1203.6064}}].

\bibitem{Kos:2014bka}
F.~Kos, D.~Poland, and D.~Simmons-Duffin, {\it {Bootstrapping Mixed Correlators
  in the 3D Ising Model}},  {\em JHEP} {\bf 1411} (2014) 109,
  [\href{http://arxiv.org/abs/1406.4858}{{\tt arXiv:1406.4858}}].

\bibitem{Fitzpatrick2013}
A.~L. Fitzpatrick, J.~Kaplan, D.~Poland, and D.~Simmons-Duffin, {\it {The
  Analytic Bootstrap and AdS Superhorizon Locality}},  {\em JHEP} {\bf 12}
  (2013) 004, [\href{http://arxiv.org/abs/1212.3616}{{\tt arXiv:1212.3616}}].

\bibitem{Komargodski2013}
Z.~Komargodski and A.~Zhiboedov, {\it {Convexity and Liberation at Large
  Spin}},  {\em JHEP} {\bf 11} (2013) 140,
  [\href{http://arxiv.org/abs/1212.4103}{{\tt arXiv:1212.4103}}].

\bibitem{Hartman2015}
T.~Hartman, S.~Jain, and S.~Kundu, {\it {Causality Constraints in Conformal
  Field Theory}},  \href{http://arxiv.org/abs/1509.00014}{{\tt
  arXiv:1509.00014}}.

\bibitem{Hartman2016}
T.~Hartman, S.~Jain, and S.~Kundu, {\it {A New Spin on Causality Constraints}},
   \href{http://arxiv.org/abs/1601.07904}{{\tt arXiv:1601.07904}}.

\bibitem{Hofman2016}
D.~M. Hofman, D.~Li, D.~Meltzer, D.~Poland, and F.~Rejon-Barrera, {\it {A Proof
  of the Conformal Collider Bounds}},
  \href{http://arxiv.org/abs/1603.03771}{{\tt arXiv:1603.03771}}.

\bibitem{paper2}
M.~F. Paulos, J.~Penedones, J.~Toledo, B.~C. van Rees, and P.~Vieira, {\it {The
  S-matrix Bootstrap II: Two Dimensional Amplitudes}},
  \href{http://arxiv.org/abs/1607.06110}{{\tt arXiv:1607.06110}}.

\bibitem{Callan1990}
C.~G. Callan and F.~Wilczek, {\it Infrared behavior at negative curvature},
  {\em Nuclear Physics B} {\bf 340} (1990), no.~2 366--386.

\bibitem{Aharony:2015zea}
O.~Aharony, M.~Berkooz, and S.-J. Rey, {\it {Rigid holography and
  six-dimensional N=(2,0) theories on AdS5 x S1}},  {\em JHEP} {\bf 03} (2015)
  121, [\href{http://arxiv.org/abs/1501.02904}{{\tt arXiv:1501.02904}}].

\bibitem{Mack2009}
G.~Mack, {\it {D-independent representation of Conformal Field Theories in D
  dimensions via transformation to auxiliary Dual Resonance Models. Scalar
  amplitudes}},  \href{http://arxiv.org/abs/0907.2407}{{\tt arXiv:0907.2407}}.

\bibitem{Mack2009a}
G.~Mack, {\it {D-dimensional Conformal Field Theories with anomalous dimensions
  as Dual Resonance Models}},  {\em Bulg. J. Phys.} {\bf 36} (2009) 214--226,
  [\href{http://arxiv.org/abs/0909.1024}{{\tt arXiv:0909.1024}}].

\bibitem{Penedones:2010ue}
J.~Penedones, {\it {Writing CFT correlation functions as AdS scattering
  amplitudes}},  {\em JHEP} {\bf 03} (2011) 025,
  [\href{http://arxiv.org/abs/1011.1485}{{\tt arXiv:1011.1485}}].

\bibitem{Fitzpatrick:2011hu}
A.~L. Fitzpatrick and J.~Kaplan, {\it {Analyticity and the Holographic
  S-Matrix}},  {\em JHEP} {\bf 10} (2012) 127,
  [\href{http://arxiv.org/abs/1111.6972}{{\tt arXiv:1111.6972}}].

\bibitem{Fitzpatrick:2011ia}
A.~L. Fitzpatrick, J.~Kaplan, J.~Penedones, S.~Raju, and B.~C. van Rees, {\it
  {A Natural Language for AdS/CFT Correlators}},  {\em JHEP} {\bf 11} (2011)
  095, [\href{http://arxiv.org/abs/1107.1499}{{\tt arXiv:1107.1499}}].

\bibitem{Paulos:2011ie}
M.~F. Paulos, {\it {Towards Feynman rules for Mellin amplitudes}},  {\em JHEP}
  {\bf 10} (2011) 074, [\href{http://arxiv.org/abs/1107.1504}{{\tt
  arXiv:1107.1504}}].

\bibitem{Freedman:1998tz}
D.~Z. Freedman, S.~D. Mathur, A.~Matusis, and L.~Rastelli, {\it {Correlation
  functions in the CFT(d) / AdS(d+1) correspondence}},  {\em Nucl. Phys.} {\bf
  B546} (1999) 96--118, [\href{http://arxiv.org/abs/hep-th/9804058}{{\tt
  hep-th/9804058}}].

\bibitem{Bargheer:2013faa}
T.~Bargheer, J.~A. Minahan, and R.~Pereira, {\it {Computing Three-Point
  Functions for Short Operators}},  {\em JHEP} {\bf 03} (2014) 096,
  [\href{http://arxiv.org/abs/1311.7461}{{\tt arXiv:1311.7461}}].

\bibitem{Minahan:2014usa}
J.~A. Minahan and R.~Pereira, {\it {Three-point correlators from string
  amplitudes: Mixing and Regge spins}},  {\em JHEP} {\bf 04} (2015) 134,
  [\href{http://arxiv.org/abs/1410.4746}{{\tt arXiv:1410.4746}}].

\bibitem{Cornalba:2007zb}
L.~Cornalba, M.~S. Costa, and J.~Penedones, {\it {Eikonal approximation in
  AdS/CFT: Resumming the gravitational loop expansion}},  {\em JHEP} {\bf 09}
  (2007) 037, [\href{http://arxiv.org/abs/0707.0120}{{\tt arXiv:0707.0120}}].

\bibitem{Heemskerk:2009pn}
I.~Heemskerk, J.~Penedones, J.~Polchinski, and J.~Sully, {\it {Holography from
  Conformal Field Theory}},  {\em JHEP} {\bf 0910} (2009) 079,
  [\href{http://arxiv.org/abs/0907.0151}{{\tt arXiv:0907.0151}}].

\bibitem{Fitzpatrick:2010zm}
A.~Fitzpatrick, E.~Katz, D.~Poland, and D.~Simmons-Duffin, {\it {Effective
  Conformal Theory and the Flat-Space Limit of AdS}},  {\em JHEP} {\bf 1107}
  (2011) 023, [\href{http://arxiv.org/abs/1007.2412}{{\tt arXiv:1007.2412}}].

\bibitem{Paulos2014a}
M.~F. Paulos, {\it {JuliBootS: a hands-on guide to the conformal bootstrap}},
  \href{http://arxiv.org/abs/1412.4127}{{\tt arXiv:1412.4127}}.

\bibitem{Simmons-Duffin2015}
D.~Simmons-Duffin, {\it {A Semidefinite Program Solver for the Conformal
  Bootstrap}},  {\em JHEP} {\bf 06} (2015) 174,
  [\href{http://arxiv.org/abs/1502.02033}{{\tt arXiv:1502.02033}}].

\bibitem{El-Showk:2016mxr}
S.~El-Showk and M.~F. Paulos, {\it {Extremal bootstrapping: go with the flow}},
   \href{http://arxiv.org/abs/1605.08087}{{\tt arXiv:1605.08087}}.

\bibitem{Poland:2010wg}
D.~Poland and D.~Simmons-Duffin, {\it {Bounds on 4D Conformal and
  Superconformal Field Theories}},  {\em JHEP} {\bf 1105} (2011) 017,
  [\href{http://arxiv.org/abs/1009.2087}{{\tt arXiv:1009.2087}}].

\bibitem{El-Showk2013}
S.~El-Showk and M.~F. Paulos, {\it {Bootstrapping Conformal Field Theories with
  the Extremal Functional Method}},  {\em Phys. Rev. Lett.} {\bf 111} (2013),
  no.~24 241601, [\href{http://arxiv.org/abs/1211.2810}{{\tt
  arXiv:1211.2810}}].

\bibitem{Beem:2015aoa}
C.~Beem, M.~Lemos, L.~Rastelli, and B.~C. van Rees, {\it {The (2, 0)
  superconformal bootstrap}},  {\em Phys. Rev.} {\bf D93} (2016), no.~2 025016,
  [\href{http://arxiv.org/abs/1507.05637}{{\tt arXiv:1507.05637}}].

\bibitem{Skenderis:2002wp}
K.~Skenderis, {\it {Lecture notes on holographic renormalization}},  {\em
  Class. Quant. Grav.} {\bf 19} (2002) 5849--5876,
  [\href{http://arxiv.org/abs/hep-th/0209067}{{\tt hep-th/0209067}}].

\bibitem{Witten:2003ya}
E.~Witten, {\it {SL(2,Z) action on three-dimensional conformal field theories
  with Abelian symmetry}},  \href{http://arxiv.org/abs/hep-th/0307041}{{\tt
  hep-th/0307041}}.

\bibitem{Aharony:2010ay}
O.~Aharony, D.~Marolf, and M.~Rangamani, {\it {Conformal field theories in
  anti-de Sitter space}},  {\em JHEP} {\bf 02} (2011) 041,
  [\href{http://arxiv.org/abs/1011.6144}{{\tt arXiv:1011.6144}}].

\bibitem{Aharony:2012jf}
O.~Aharony, M.~Berkooz, D.~Tong, and S.~Yankielowicz, {\it {Confinement in
  Anti-de Sitter Space}},  {\em JHEP} {\bf 02} (2013) 076,
  [\href{http://arxiv.org/abs/1210.5195}{{\tt arXiv:1210.5195}}].

\bibitem{Doyon:2003nb}
B.~Doyon, {\it {Two point correlation functions of scaling fields in the Dirac
  theory on the Poincare disk}},  {\em Nucl. Phys.} {\bf B675} (2003) 607--630,
  [\href{http://arxiv.org/abs/hep-th/0304190}{{\tt hep-th/0304190}}].

\bibitem{Doyon:2004fv}
B.~Doyon and P.~Fonseca, {\it {Ising field theory on a Pseudosphere}},  {\em J.
  Stat. Mech.} {\bf 0407} (2004) P07002,
  [\href{http://arxiv.org/abs/hep-th/0404136}{{\tt hep-th/0404136}}].

\bibitem{Osborn:1999az}
H.~Osborn and G.~M. Shore, {\it {Correlation functions of the energy momentum
  tensor on spaces of constant curvature}},  {\em Nucl. Phys.} {\bf B571}
  (2000) 287--357, [\href{http://arxiv.org/abs/hep-th/9909043}{{\tt
  hep-th/9909043}}].

\bibitem{Dirac1936}
P.~A. Dirac, {\it Wave equations in conformal space},  {\em Annals of
  Mathematics} (1936) 429--442.

\bibitem{Costa:2016hju}
M.~S. Costa, T.~Hansen, J.~Penedones, and E.~Trevisani, {\it {Projectors and
  seed conformal blocks for traceless mixed-symmetry tensors}},  {\em JHEP}
  {\bf 07} (2016) 018, [\href{http://arxiv.org/abs/1603.05551}{{\tt
  arXiv:1603.05551}}].

\bibitem{Goncalves:2014rfa}
V.~Goncalves, J.~Penedones, and E.~Trevisani, {\it {Factorization of Mellin
  amplitudes}},  {\em JHEP} {\bf 10} (2015) 040,
  [\href{http://arxiv.org/abs/1410.4185}{{\tt arXiv:1410.4185}}].

\bibitem{Costa:2012cb}
M.~S. Costa, V.~Goncalves, and J.~Penedones, {\it {Conformal Regge theory}},
  {\em JHEP} {\bf 12} (2012) 091, [\href{http://arxiv.org/abs/1209.4355}{{\tt
  arXiv:1209.4355}}].

\bibitem{Fitzpatrick:2011dm}
A.~L. Fitzpatrick and J.~Kaplan, {\it {Unitarity and the Holographic
  S-Matrix}},  {\em JHEP} {\bf 10} (2012) 032,
  [\href{http://arxiv.org/abs/1112.4845}{{\tt arXiv:1112.4845}}].

\bibitem{Pappadopulo:2012jk}
D.~Pappadopulo, S.~Rychkov, J.~Espin, and R.~Rattazzi, {\it {OPE Convergence in
  Conformal Field Theory}},  \href{http://arxiv.org/abs/1208.6449}{{\tt
  arXiv:1208.6449}}.

\bibitem{Kim:2015oca}
H.~Kim, P.~Kravchuk, and H.~Ooguri, {\it {Reflections on Conformal Spectra}},
  \href{http://arxiv.org/abs/1510.08772}{{\tt arXiv:1510.08772}}.

\end{thebibliography}\endgroup
\bibliographystyle{JHEP}

\end{document}